\documentclass[fleqn,usenatbib]{mnras}

\usepackage{newtxtext,newtxmath}

\usepackage[T1]{fontenc}

\DeclareRobustCommand{\VAN}[3]{#2}
\let\VANthebibliography\thebibliography
\def\thebibliography{\DeclareRobustCommand{\VAN}[3]{##3}\VANthebibliography}

\usepackage{graphicx}	
\usepackage{amsmath}	

\title[GCs of Perseus LSB Galaxies]{The PIPER Survey. II. The Globular Cluster
Systems of Low Surface Brightness Galaxies in the Perseus Cluster}

\author[S. R. Janssens et al.]{
    Steven~R.~Janssens$^{1}$\thanks{E-mail: steven.janssens@utoronto.ca},
    Duncan~A.~Forbes$^{1}$\thanks{E-mail: dforbes@swin.edu.au},
    Aaron~J.~Romanowsky$^{2,3,4}$,
    Jonah~Gannon$^{1}$,
    Joel~Pfeffer$^{1}$,
    \newauthor{
        Warrick~J.~Couch$^{1}$,
        Jean~P.~Brodie$^{1}$,
        William~E.~Harris$^{5}$,
        Patrick~R.~Durrell$^{6}$,
        Kenji~Bekki$^{7}$
    }
\\
$^{1}$Centre for Astrophysics and Supercomputing, Swinburne University, Hawthorn VIC 3122, Australia\\
$^{2}$Department of Physics \& Astronomy, San Jos\'{e} State University, One Washington Square, San Jose, CA 95192, USA\\
$^{3}$Department of Astronomy \& Astrophysics, University of California, Santa Cruz, CA 95064, USA\\
$^{4}$University of California Observatories, 1156 High Street, Santa Cruz, CA 95064, USA\\
$^{5}$Department of Physics \& Astronomy, McMaster University, 1280 Main Street West, Hamilton, L8S 4M1, Canada\\
$^{6}$Department of Physics, Astronomy, Geology and Environmental Sciences,
Youngstown State University, Youngstown OH 44555, USA\\
$^{7}$ICRAR, The University of Western Australia, 35 Stirling Hwy, Crawley WA 6009, Australia\\
}

\date{Accepted XXX. Received YYY; in original form ZZZ}

\pubyear{2024}

\begin{document}
\label{firstpage}
\pagerange{\pageref{firstpage}--\pageref{lastpage}}
\maketitle

\begin{abstract}
    We present \textit{Hubble Space Telescope} ACS/WFC and WFC3/UVIS imaging for a sample of 50 low surface brightness (LSB) galaxies in 
    the $\sim$10$^{15}$ M$_{\odot}$ Perseus cluster, which were 
    originally identified in ground-based imaging. 
    We measure the structural properties of these  galaxies and estimate the total number of globular clusters (GCs) they host. Around half of our sample galaxies meet the strict definition of an ultra-diffuse galaxy (UDG), while the others are UDG-like but are either somewhat more compact or slightly brighter. A small number of  galaxies reveal 
    systems with many tens of GCs, 
    rivalling some of the richest GC systems known around UDGs in the Coma cluster.
    We find the sizes of rich GC systems, in terms of their half-number radii, extending to $\sim$1.2 times the half-light radii of their host  galaxy on average.
    The mean colours of the GC systems are the same, within the uncertainties, as those of their host galaxy stars. This suggests that GCs and galaxy field stars may have formed at the same epoch from the same enriched gas. It may also indicate a significant contribution from disrupted GCs to the stellar component of the host galaxy as might be expected in the `failed galaxy' formation scenario for UDGs. 
    
\end{abstract}

\begin{keywords}
galaxies: dwarf -- galaxies: star clusters: general -- galaxies: formation.
\end{keywords}



\section{Introduction}

Ultra-diffuse galaxies (UDGs) are a class of extended low surface brightness
(LSB)
galaxies.
Their abundance in galaxy clusters was discovered with the Dragonfly Telephoto
Array where a large population of galaxies with effective radii $R_\mathrm{e}
\gtrsim 1.5~\mathrm{kpc}$ and central surface brightnesses $\mu_{0{,}g}
\gtrsim 24~\mathrm{mag}~\mathrm{arcsec}^{-2}$ were identified in the Coma cluster \citep{vandokkum2015}.

High resolution imaging of Coma cluster UDGs with the {\it Hubble Space
Telescope} ({\it HST}) revealed some to have remarkably rich globular cluster
(GC) systems for galaxies with stellar masses of around $10^8~M_{\odot}$
\citep{vandokkum2017,lim2018,amorisco2018}. 
This was further reinforced by \cite{forbes2020} who combined the GC counts
from these three studies.
While some Coma UDGs were consistent with hosting no GCs, others hosted up to
a hundred GCs.
For example, Dragonfly (DF) 44 was estimated by \cite{vandokkum2017} to have
$76 \pm 18$ GCs.
However, also using {\it HST} imaging, \cite{2022MNRAS.511.4633S} examined six
Coma UDGs and found much lower counts than \cite{vandokkum2017} or
\cite{lim2018}.
For DF44, \cite{2022MNRAS.511.4633S} suggested a total GC system of
$20^{+6}_{-5}$. 
\cite{forbes2024} have shown that the lower GC counts of
\cite{2022MNRAS.511.4633S} are driven by their measured GC system size
(half-number radius) relative to the galaxy half-light radius.
While \cite{forbes2024} favoured the numbers found by \cite{vandokkum2017}
and \cite{lim2018}, without spectra of GC candidates it is unlikely this
matter will be fully settled in the near future.

The GC systems of UDGs in a number of other clusters have also been studied. These include Fornax \citep{2019MNRAS.484.4865P}, Virgo \citep{2020ApJ...899...69L} and Hydra~I \citep{2022A&A...665A.105L}, but few, if any, UDGs possess the rich GC systems found in the Coma cluster. This, along with the claims of \cite{2022MNRAS.511.4633S}, have led some to question the reliability of the rich GC systems in Coma UDGs. Another possibility is that only the most massive clusters like Coma host a population of UDGs with rich GC systems. Clearly an equivalent {\it HST} study of a cluster with a total mass similar to Coma ($0.5 < M_{200} < 1.25 \times 10^{15}~M_{\odot}$; \citealt{2022NatAs...6..936H}) is required. 

As well as determining the ratio of GC system to galaxy size, another important measurement is the mean colour of the GC system and how it compares with the host galaxy light. If galaxy stars formed independently of the GCs and over a longer period, then we might expect the galaxy to be on average redder than the GCs. 
However, a similar colour would suggest a similar age and metallicity for GCs
and galaxy stars, indicating they may have formed from the same enriched gas
at the same epoch. Such stellar populations might be expected in the `failed
galaxy' formation scenario whereby GCs form and, by some mechanism, the host
galaxy is quenched at early times \citep{danieli2022, forbes2024}. 
Disruption of GCs over time will contribute field stars to the host galaxy. If the stars from disrupted GCs dominate the luminosity of the galaxy, then similar mean colours in the remaining GCs and the galaxy stars would be measured. 
In a recent detailed analysis of the colours of GCs in two UDGs (DF2 and DF4 in the NGC 1052 group) by 
\cite{vandokkum2022}, they found that in both cases the GCs are slightly redder on average than their host galaxy. 

We present multi-band {\it HST} imaging of the Perseus cluster obtained as
part of the Program for Imaging of the PERseus cluster of galaxies (PIPER)
survey \citep[hereafter Paper~I]{harris2020}.
The large scale structure of GCs and
their spatial associations on the sky have been previously examined by Paper~I
and \cite{2022ApJ...935....3L}.
Here we measure galaxy properties using {\it HST} for a sample of 50 LSB
galaxies originally identified in ground-based imaging. Measured properties
include  their  magnitudes, mean colours, half-light radii and ellipticities.
We also identify their GC candidates and, after correcting for incompleteness
and  contamination, estimate their total GC numbers. We measure the
half-number radii and mean colours of the GC systems, and compare them to the
host galaxy.
We follow Paper~I and adopt a distance $d = 75~\mathrm{Mpc}$ to the
Perseus cluster, which corresponds to a distance modulus $\mu = 34.38$.
Perseus has a virial radius $R_{200} = 1.79 \pm 0.04~\mathrm{Mpc}$ and a total
mass $M_{200} = 6.65^{+0.43}_{-0.46} \times 10^{14}~M_\odot$ based on X-ray
observations of the intracluster medium \citep{simionescu2011}.
\cite{aguerri2020} measured a larger, more massive cluster with $R_{200} =
2.2~\mathrm{Mpc}$ and $M_{200} = 1.2 \times 10^{15}~M_\odot$ from
spectroscopic measurements of 403 galaxies in Perseus.
Thus the Perseus cluster has a mass similar to that of the Coma cluster (whose UDG population has been well-studied).
All photometry is in the VEGAmag system.
Galactic extinction is quite high in the direction of the Perseus cluster with
up to $A_V \sim 0.5$.
This means the galaxies and GCs are effectively as faint as in the Coma cluster, but they are still slightly better resolved.
Here we apply
corrections from the \cite{schlafly2011} extinction maps to all colours and magnitudes.\footnote{Using the online
calculator at \url{https://ned.ipac.caltech.edu/forms/calculator.html}.
The maps have a resolution of a few arcmin, while deep imaging of Perseus from Subaru Hyper Suprime-Cam \citep{2022MNRAS.510..946G} shows strong cirrus variations on scales of a few arcsec.
This effect will introduce additional scatter in our photometric measurements
(particularly for galaxies R16, R23, R79, W1, W5, W8, W22, W28, W29).}

\section{Data and Methods}

\subsection{\textit{HST} Observations}

The Program for Imaging of the 
PERseus cluster of galaxies (PIPER) used
ACS/WFC and WFC3/UVIS aboard \textit{HST} in parallel to observe ten pairs of fields
outside of the Perseus cluster core and 
five focused on NGC~1275 and other giant cluster galaxies in the core (see
Figure~1 of Paper~I for the arrangement of the fields).
The field pairs, or ``visits'', were chosen to contain 50 LSB galaxies.
They were selected from the catalogue of Perseus LSB galaxies by
\cite{2017MNRAS.470.1512W} (33 targets, designated W) and supplemented by
additional LSB targets found through visual inspection of archival
$G^{\prime}$- and $R^{\prime}$-band CFHT/MegaCam imaging\footnote{The MegaPipe
\citep{gwyn2008} processed images are available from the Canadian Astronomy
Data Centre with \texttt{observationID} \texttt{MegaPipe.075.263}.} (17
targets, designated R), whose UDG-like nature was confirmed using GALFIT
\citep{peng2002}.
The target selection prioritized the largest, UDG-like galaxies ($R_{\rm e}
\gtrsim$~2 kpc) while also including as many LSB galaxies as possible in each
visit.
The five core fields cover the inner ${\sim}230~\mathrm{kpc}$
(${\sim}3~\mathrm{arcmin}$) around NGC~1275.
The outer fields range from a minimum distance of ${\sim}240~\mathrm{kpc}$
(${\sim}11~\mathrm{arcmin}$) from NGC~1275 out to a maximum of
${\sim}780~\mathrm{kpc}$ (${\sim}36~\mathrm{arcmin}$), with a mean distance of
${\sim}475~\mathrm{kpc}$ (${\sim}22~\mathrm{arcmin}$), though the sampling is
fairly uniform and there is a strong bias to the West side of the cluster.
One target, W27, turned out to not exist at the given coordinates in the 
\cite{2017MNRAS.470.1512W} catalogue. 

Each visit was observed for two orbits.
For ACS, one orbit used the F814W filter and the other F475W.
For WFC3, F814W and F475X were the two filters used.
The calibrated, flat fielded and charge transfer efficiency (CTE) corrected
individual exposures (\texttt{flc} files) were obtained from MAST.
For each exposure, we ensured that either a GAIA DR2 or DR3 \textit{a
posteriori} WCS solution contained in an extension headerlet was applied.
These were combined using \textsc{AstroDrizzle} to a final pixel scale of
$0.03\arcsec$, and were weighted with an inverse-variance weight map.
Exposure time maps were obtained by running \textsc{AstroDrizzle} a
second time, and weighting by exposure time, with the other products being discarded.
We were notified of guiding issues potentially affecting the visits to
PERSEUS-UDG02, PERSEUS-UDG06 and NGC1275-F1 (see Table~1 of Paper~I for
details of all the visits).
However, only a single WFC3 F475X exposure for the PERSEUS-UDG02 visit
exhibited streaks indicative of a loss of guiding.
The affected \texttt{flc} file was removed before combining with
\textsc{AstroDrizzle}.
With only two remaining exposures, this image is more contaminated by cosmic rays
compared to the others.

\subsection{Photometry}\label{sec:photometry}

\textsc{SExtractor} \citep{bertin1996} was run in dual-image mode on each
image pair (F475W/F814W for ACS, F475X/F814W for WFC3) using the F814W image
as the detection image.
Magnitudes were measured in 5 pixel diameter apertures.
Since the purpose of this catalog is to detect unresolved or marginally
resolved GCs and not galaxies, a very small background cell size of
$\mathtt{BACK\_SIZE} = 24$ pixels was used.
This both aids in the detection of GCs by subtracting off large scale
structure, including the stellar bodies of galaxies, as well as provides a
more accurate background measurement.
As a final aid in ensuring our GC detection is complete around our target
galaxies, we subtracted off the best-fit \texttt{Imfit} \citep{erwin2015} model
(described in \S \ref{sec:imfit}) before running \textsc{SExtractor}.
All magnitudes were placed on the VEGAmag system\footnote{VEGAmag was chosen
over AB as the Paper~I $\mathrm{F475X}_\mathrm{WFC3}$ to
$\mathrm{F475W}_\mathrm{ACS}$ transformation is defined in the Vega system.}.
\textsc{ACStools} \citep{lim2020} was used to retrieve the VEGAmag zeropoints
for the ACS images, and \textsc{stsynphot} \citep{stsci2020} was used to
calculate them for the WFC3 images.
Galactic extinction values \citep{schlafly2011} vary by several tenths of a magnitude from field
to field. Here we used the extinction value for the centre of each field and applied that to all
sources within. 

Aperture corrections were then applied.
For the ACS images, we first corrected from a 5 pixel diameter aperture to a
$1\arcsec$ diameter aperture by computing the encircled energy using a DAOPHOT
model point spread function (PSF; see \S \ref{sec:artstars}) for each image.
We then corrected from the $1\arcsec$ diameter aperture to
infinity using table 5 in \cite{sirianni2005}\footnote{The difference between
the \cite{sirianni2005} and the more recent \cite{bohlin2016} ACS aperture
corrections are negligible, at 0.006 mag and 0.01 mag for F475W and F814W,
respectively, and have no impact on the results.}.
We performed the same procedure for the WFC3/UVIS magnitudes, but instead used the
UVIS2 encircled energy
fractions\footnote{\url{https://www.stsci.edu/hst/instrumentation/wfc3/data-analysis/photometric-calibration/uvis-encircled-energy}} to correct from a
$1\arcsec$ diameter aperture to infinity.
For ACS, the total 5 pixel to infinity diameter aperture correction values
range from 0.79 to 0.94 mag for F814W and 0.67--0.88 mag for F475W, with the
range due to the PSF variation between \textit{HST} visits.
For WFC3, the total aperture correction values are 0.67--0.80 mag for F814W
and 0.59--0.87 mag for F475X.

We placed all photometry on the ACS system.
\cite{deustua2018} found $\mathrm{F814W}_\mathrm{WFC3} \simeq
\mathrm{F814W}_\mathrm{ACS}$.
For the WFC3 fields, the blue filter is F475X.
Following Paper~I, we transformed F475X to $\mathrm{F475W}_\mathrm{ACS}$
using
\begin{equation}\label{eqn:f475w}
\mathrm{F475W}_\mathrm{ACS} = \mathrm{F475X}_\mathrm{WFC3} +
    (0.16 \pm 0.02)(\mathrm{F475X} - \mathrm{F814W})_\mathrm{WFC3}.
\end{equation}
There is a small region of overlap between the ACS and WFC3 imaging 
on
the outskirts of NGC~1275 yielding 133 point sources with both ACS and WFC3
photometry, the vast majority of which have colours and magnitudes consistent
with GCs.
The distribution of WFC3 derived $(\mathrm{F475W} -
\mathrm{F814W})_\mathrm{ACS}$ colours agrees well with the native ACS colours
for these objects; the mean colour difference is 0.00 with a standard
deviation of 0.27.

\subsection{Artificial Star Tests and Completeness Limits}\label{sec:artstars}

DAOPHOT \citep{stetson1987} was used to construct a model PSF
for each \textit{HST} ACS and WFC3 image in each filter.
The exposure times in the headers were used to convert the images from
units of electrons per second to electrons, which is necessary for DAOPHOT to
understand the noise.
We followed the standard DAOPHOT procedure of running the \texttt{FIND},
\texttt{PHOT} and \texttt{PICK} commands to select suitable PSF stars in each
image.
These were run without any user intervention.
We then culled the list of PSF stars created by \texttt{PICK} by cross matching
it to a list of suitable bright stars selected \textit{a priori} from the
\textsc{SExtractor} catalog.
Suitable PSF stars have $\mathtt{FLAGS} = 0$ (i.e.\ no photometry issues with
this object), $18 < \mathtt{MAG\_AUTO} < 23.5$, $1.25~\mathrm{pixels} <
\mathtt{FLUX\_RADIUS} < 3.25~\mathrm{pixels}$, no bad or empty pixels within
the DAOPHOT fitting radius ($r = 10~\mathrm{pixels}$) and no neighbours within the DAOPHOT
PSF radius ($r = 51~\mathrm{pixels}$).
The number of PSF stars varies from 25 stars in the NGC1275-F5 WFC3 F475X
image to over 200 in the NGC1275-F1 ACS F814W image.
We adopt DAOPHOT defaults in defining the PSF model, namely a Gaussian
analytic profile along with an empirical lookup table of corrections.
Given the small expected variation of the PSF across a given field, no spatial variation of the PSF was
applied.
We then used the \texttt{ADDSTAR} routine in \texttt{DAOPHOT} to create an
image of the PSF model to use for all subsequent tasks.

Artificial star tests were then performed by scaling the PSF images to the
desired total magnitude and adding them to our images.
We selected 5000 random positions in each F814W and F475W/F475X image pair.
The \textsc{SExtractor} segmentation maps were used to ensure a selected
location was empty.
Note that due to the small \textsc{SExtractor} background cell size used, most
of the pixels belonging to UDGs and other galaxies are empty in the
segmentation maps and thus are valid regions to inject artificial stars.
The F814W total magnitudes of the artificial stars were drawn uniformly
between $22.5 < \mathrm{F814W} < 29.5$.
An $\mathrm{F475W} - \mathrm{F814W}$ colour was then drawn from a skewnormal
distribution with mean $\mu=0.7$, standard deviation $\sigma=1.6$ and skewness
parameter $\alpha=10$ in order to roughly match the observed colour range and
distribution, with most sources (and GCs) having $\mathrm{F475W} -
\mathrm{F814W} \lesssim 2$ (Paper~I).
When injecting sources into a WFC3 image, the F475W magnitude was transformed to F475X by
inverting Equation~\ref{eqn:f475w}.
Finally, Galactic extinction was added back on to the injected magnitudes.

The injected images were then processed in an identical manner to the real
images.
The artificial sources were recovered by matching the nearest detection within
a maximum distance of 3 pixels ($0.09\arcsec$) from the expected position,
with no other restrictions on a match.
Figure \ref{fig:completeness_curves} shows the recovered fraction of
artificial stars in F814W with bins of 0.25 magnitudes.
The completeness curves were fitted using
\begin{equation}\label{eqn:fm}
f(m) = \frac{1}{1 + e^{\alpha(m - m_{50})}},
\end{equation}
where $m_{50}$ is the magnitude at which the recovered fraction has dropped to
50\% and $\alpha$ controls how steeply the completeness drops off
\citep{harris2016}.
The observed depths between visits for each camera are very similar,
with the observed variation simply reflecting the extinction variations in
this part of the sky.
The dotted lines show the completeness level at the expected GCLF turnover
magnitude $m_\mathrm{TO{,}F814W} = 26.3$, corresponding to
$M_{\mathrm{TO}{,}I} = -8.1$ \citep{2007ApJ...670.1074M} at a distance of 75~Mpc.
The data are between 71\% and 97\% complete at the turnover magnitude in every field, and at least 80\% complete in all UDG fields,
with an average completeness at $m_\mathrm{TO{,}F814W}$ of 87\%.

\subsection{Point Source Selection}

\begin{figure*}
    \includegraphics[width=0.99\textwidth]{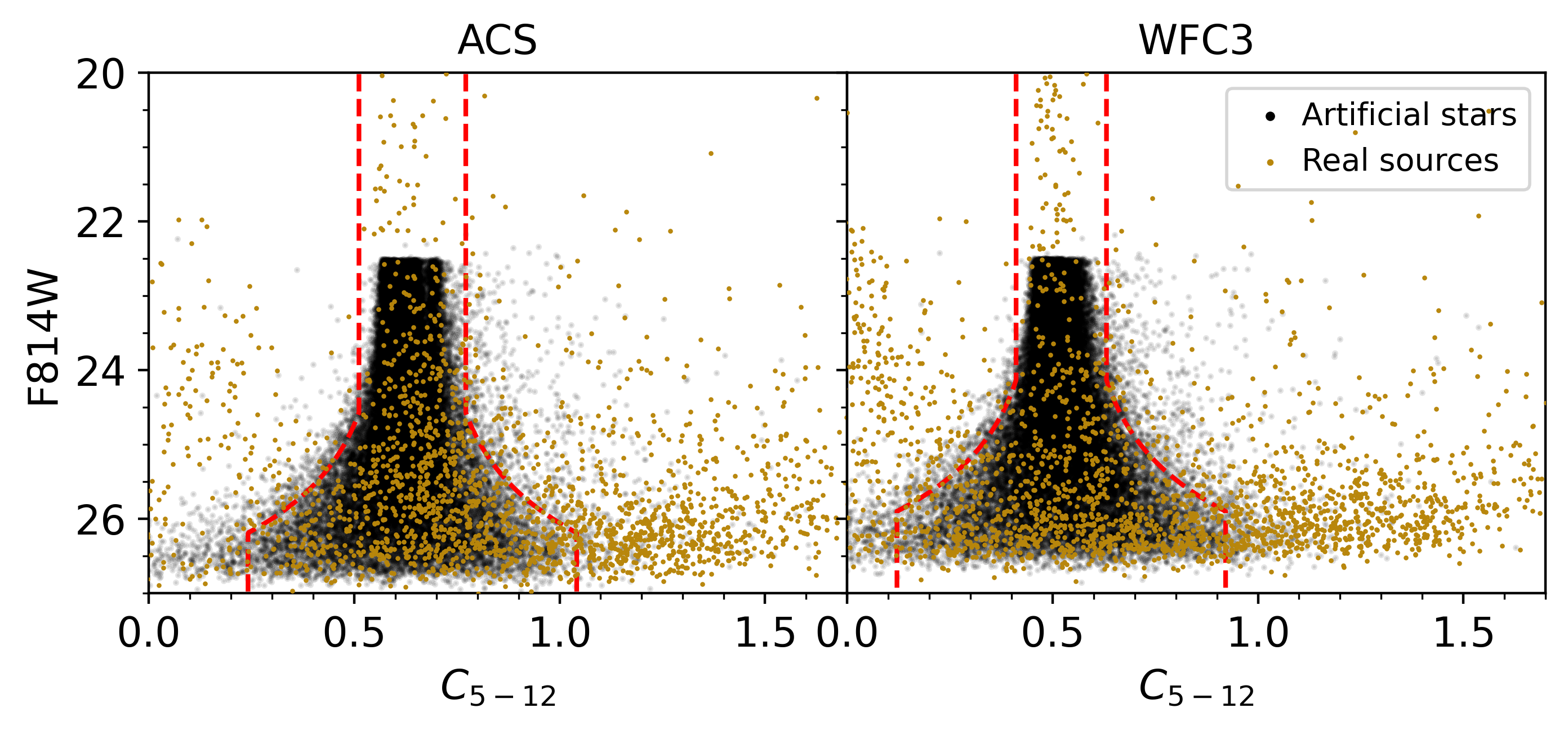}
	\caption{
    Concentration based selection of point sources.
    $C_{5-12}$ is the difference in F814W magnitude between 5 and 12 pixel
    diameter apertures.
    Since the PSF differs slightly between the ACS/WFC and WFC3/UVIS cameras,
    a selection for each is made based on the observed locus of artificial
    stars.
    The black points are our artificial stars and the gold points show a
    representative subset of our real sources.
    The dashed red curves are our point source selection boundaries, which encompass
    the bulk of our artificial stars.
    See text for details.
    \label{fig:concentration}
	}
\end{figure*}

Figure \ref{fig:concentration} shows our selection of point sources based on
their concentration $C_{5-12}$, the difference in F814W magnitude measured in
5 and 12 pixel diameter apertures.
Sources from all fields are shown. In black are all our artificial stars, and
in gold are a representative subset of real sources from the images.
Since the PSF is slightly different between the two cameras, a slightly
different selection has to be made for each camera.
The dashed vertical red curves are our adopted magnitude dependent point source
selection that encompasses most of our artificial stars.
The selection broadens from a minimum of $\pm0.13$ from the mean concentration
value for ACS ($\pm0.11$ for WFC3) to a maximum of $\pm0.40$ for both cameras.
These extremes were estimated from the brightest and faintest artificial
stars, with the value at the faint end chosen as a compromise between
encompassing as many artificial stars as possible and minimizing the
number of contaminants.
Between these extremes we fitted the observed standard deviation of $C_{5-12}$ of
artificial stars as a function of magnitude with an exponential function and
selected all detections within $\pm2\sigma$.

To check that GCs at 75~Mpc would pass these concentration cuts, we used
GALFIT \citep{peng2002} to create \cite{king1962} profiles convolved with our
PSFs and measured the resulting $C_{5-12}$.
We simulated objects with core radii between 0.5 and 10 pc and $1.5 <
\log(r_t/r_c) < 2.5$, where $r_t$ and $r_c$ are the tidal and core radii,
respectively, thus spanning GCs to ultra-compact dwarfs
\citep[UCDs;][]{evstigneeva2008}.
Half-light radii $r_h$ were measured on a separate high-resolution model
without PSF convolution.
Objects with $r_h \lesssim 5~\mathrm{pc}$ pass our point source cuts, which
matches the expected resolution limit of \textit{HST} at 75~Mpc (we refer the
reader to section 4.1 of Paper~I for a discussion on the size distribution of
intracluster GC candidates in a handful of the outer fields).
A clean point source selection with fewer background galaxy contaminants comes
at the cost of possibly missing a fraction of GCs with larger radii that are
partially resolved.
However, we believe this fraction to be low as inspection of sources around
our sample galaxies reveals only a handful of compact sources that fail this
point source cut but have the magnitudes and colours of GCs (see \S
\ref{sec:gc_selection}).

\subsection{Globular Cluster Selection}\label{sec:gc_selection}

\begin{figure*}
    \includegraphics[width=0.99\textwidth]{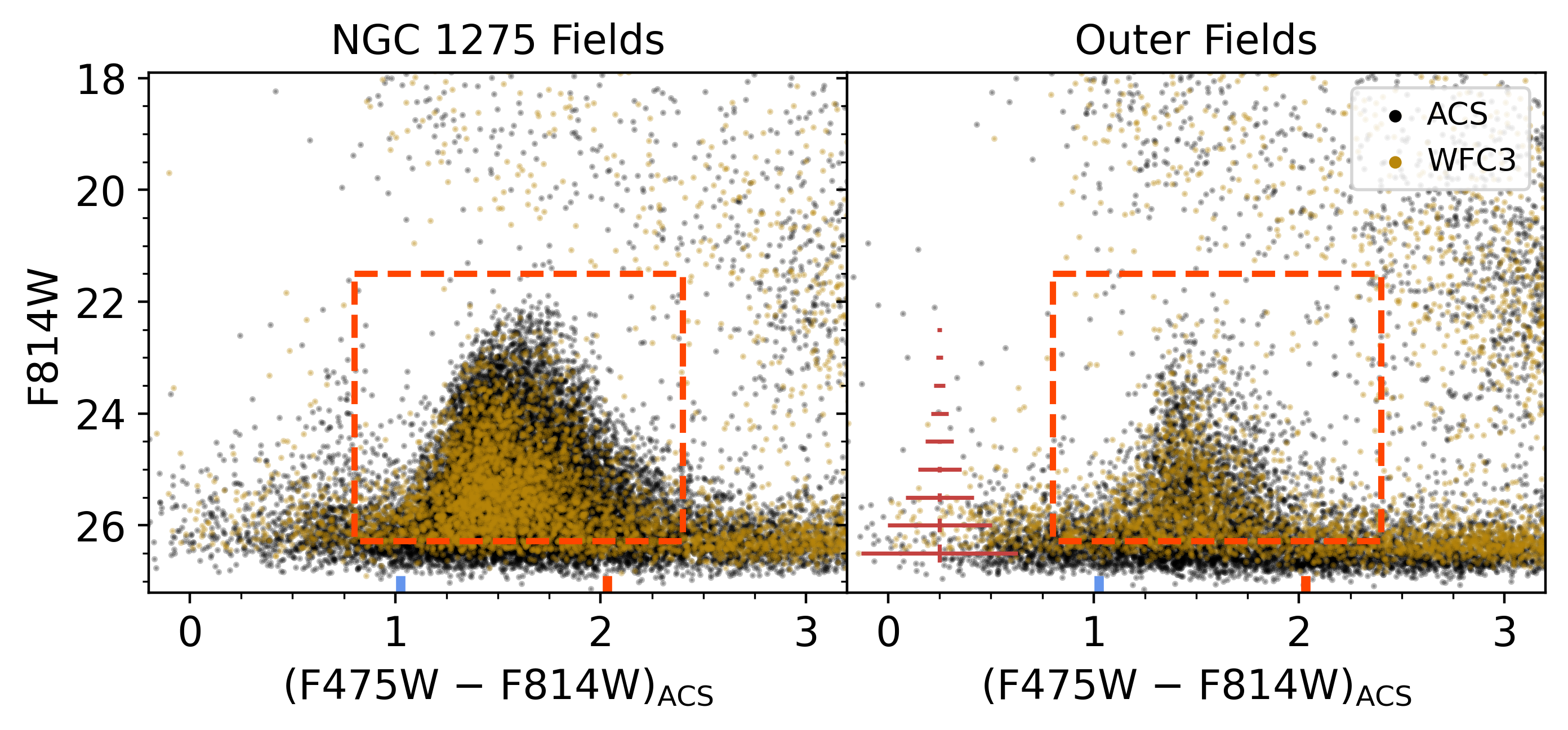}
	\caption{
    Combined ACS and WFC3 F475W/F814W colour--magnitude diagrams of point sources showing our
    selection of GCs.
    Sources from the cluster core NGC~1275 fields are shown in the left panel, while the right panel shows sources from the outer fields.
    WFC3 photometry (gold points) has been transformed to the ACS system (see main text for
    details).
    The dashed red boxes show our GC selection of $0.8 < \mathrm{F475W} -
    \mathrm{F814W} < 2.4$ and $21.5 < \mathrm{F814W} < 26.3$, the faint end of
    which is the predicted turnover magnitude of the GCLF.
    The blue and red ticks along the bottom show the predicted colours of SSPs
    with the most extreme possible GC properties. The blue tick is a 5 Gyr old
    GC with $[\mathrm{Fe}/\mathrm{H}] = -2.2$. The red tick is 13 Gyr old and
    has $[\mathrm{Fe}/\mathrm{H}] = 0.6$.
    The red error bars on the left are characteristic errors from the
    artificial star tests.
    \label{fig:cmd}
	}
\end{figure*}

At a distance of 75 Mpc, GCs in the Perseus cluster are essentially point sources when imaged by the {\it HST}. 
From point sources, GCs were selected on the basis of magnitude and colour.
Figure \ref{fig:cmd} shows the colour--magnitude diagram of all
point sources in the F475W and F814W filters. 
WFC3 sources have been transformed to the ACS filter system and are plotted in gold.
The dashed red box shows our GC selection which comprises all point sources 21.5 $<$ F814W $<$ 26.3, 
and within the colour range $0.8 <
\mathrm{F475W} - \mathrm{F814W} < 2.4$.
The faint limit corresponds to the predicted GCLF turnover magnitude. The bright magnitude cutoff corresponds to a generous stellar mass for a GC of
${\sim}10^7~M_\odot$ and so would accommodate objects classified as UCDs. However, examination of 
Figure \ref{fig:cmd} shows there are very few sources brighter than $\mathrm{F814W} \sim 22$ -- though such UCDs may have failed the concentration cuts.
The red error bars on the left-hand side show characteristic colour errors from the
artificial star tests in bins of 0.5 magnitudes, restricted to artificial
stars with a colour inside the GC selection box.
The red and blue ticks along the bottom show the predicted colours for a large possible range of properties spanned by GCs using FSPS \citep{conroy2009, conroy2010}. 
The blue tick has an age of 5 Gyr and is metal poor with
$[\mathrm{Fe}/\mathrm{H}] = -2.2$.
The red tick is 13 Gyr old and is metal rich with $[\mathrm{Fe}/\mathrm{H}] =
0.6$.
While it is unlikely an individual GC would possess these properties, this
exercise is merely to ascertain the colour range in which we expect GCs to
fall.
Our GC colour selection was expanded beyond this range to account for the large
photometric scatter at faint magnitudes.

\begin{figure*}
	\includegraphics[width=0.99\textwidth]{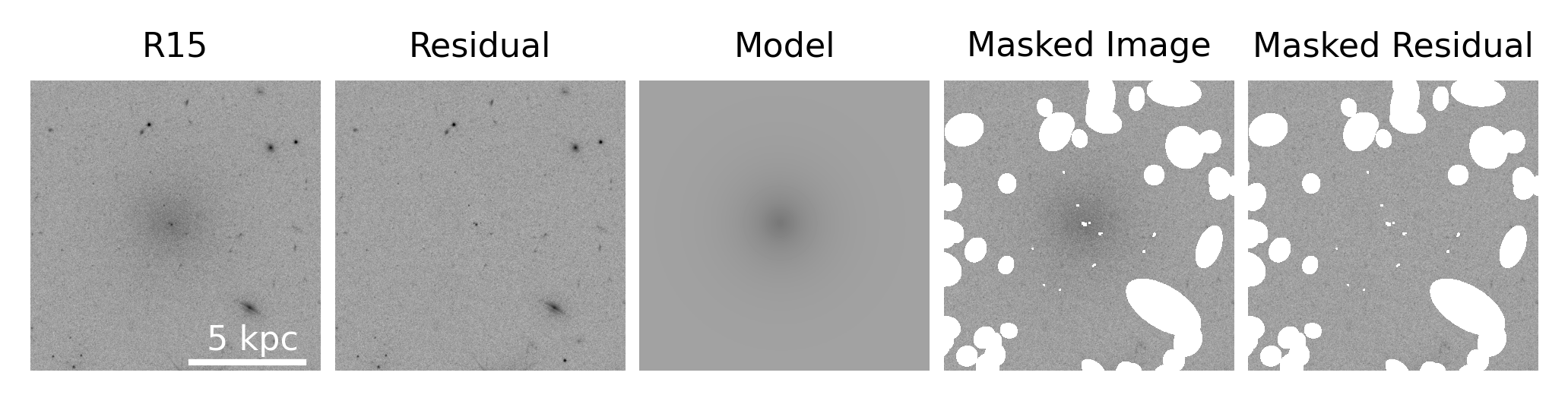}
	\includegraphics[width=0.99\textwidth]{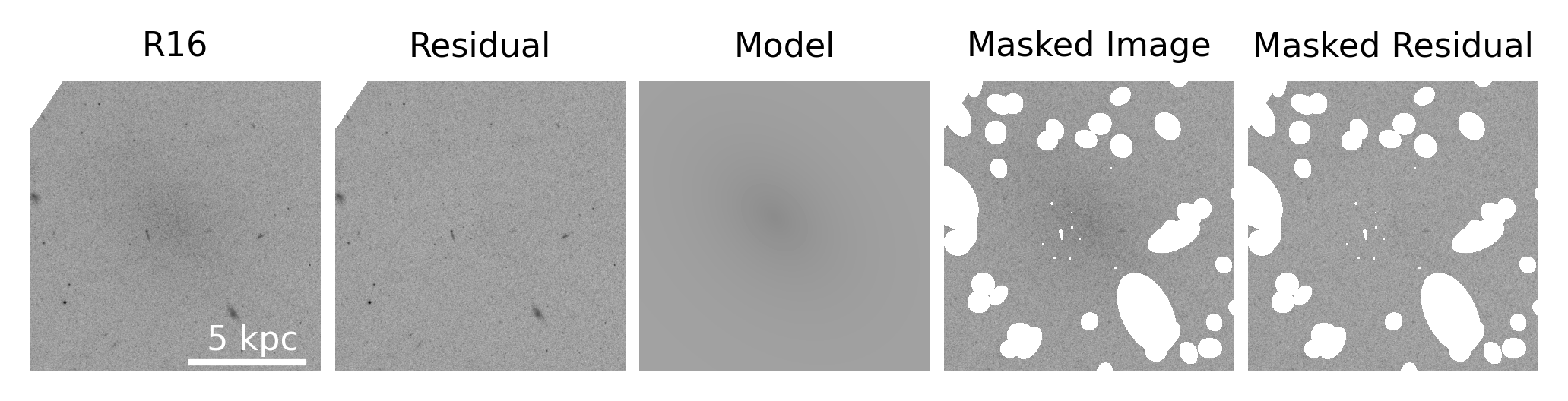}
	\includegraphics[width=0.99\textwidth]{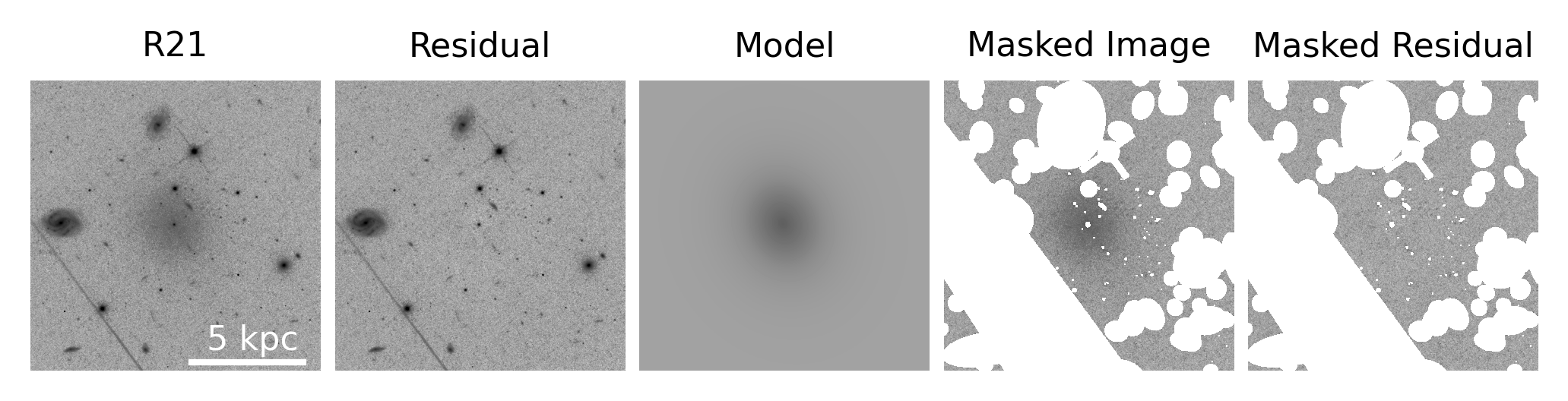}
	\includegraphics[width=0.99\textwidth]{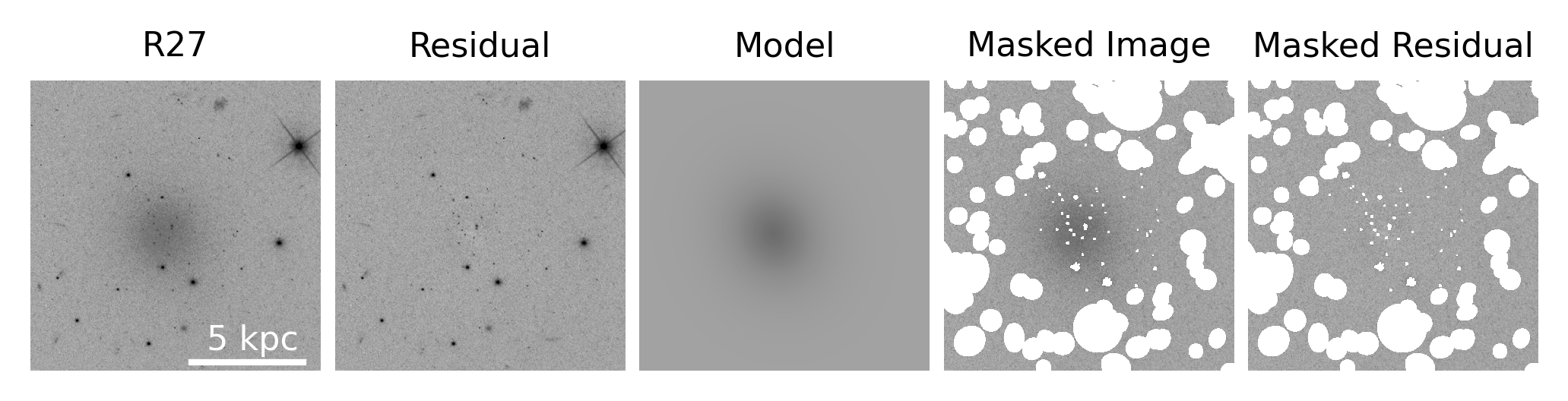}
	\includegraphics[width=0.99\textwidth]{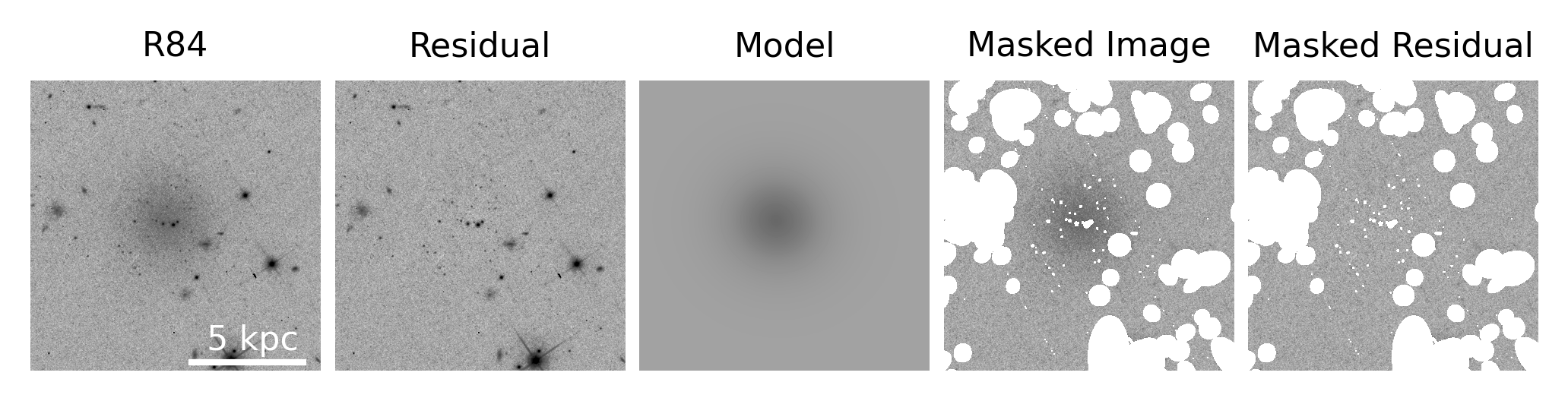}
    \caption{
    \texttt{Imfit} results for 
    five example galaxies (among the most luminous in the sample): R15, R16, R21, R27 and R84. 
    The cutouts are 13 kpc on a side (${\sim}36\arcsec$) with North oriented up and East to the left.
    For each galaxy, from left to right, is shown the
    \textit{HST}/ACS F814W image, the residual image after model subtraction,
    the \texttt{Imfit} model image, the image with mask applied, and the
    residual image with mask applied. The mask includes any central nuclei (e.g. R21). 
    \label{fig:imfit}
	}
\end{figure*}

\subsection{Galaxy Fitting}\label{sec:imfit}

We used
\texttt{Imfit}\footnote{\url{https://www.mpe.mpg.de/~erwin/code/imfit/}}
\citep{erwin2015} to fit a single component S\'{e}rsic model plus a sky
pedestal to the F814W and F475W/F475X image of each of the 50 known UDGs
targeted as part of this program.
\texttt{Imfit} was run on $150{\arcsec} \times 150{\arcsec}$ cutouts centred on each
galaxy in MCMC mode.
Since the cutouts are in units of $\mathrm{electrons}~\mathrm{s}^{-1}$ and
were sky subtracted by \texttt{AstroDrizzle}, the total exposure time and
original sky value (computed from the \texttt{MDRIZSKY} values in the
\texttt{flc} headers) were passed to \texttt{imfit-mcmc}.
Other sources in the cutouts were excluded by first masking all pixels belonging
to other objects in the \textsc{SExtractor} segmentation map.
Note that the mask includes any GCs or nuclei, preventing them -- particularly central nuclei -- from impacting the fit.
We then expanded the masks around sources by masking all pixels within a Kron
ellipse of a source with semi-major axis equal to $3 \times \mathtt{A\_IMAGE}
\times \mathtt{KRON\_RADIUS}$.
The semi-major axis of the mask ellipse was not permitted to be less than 20
pixels ($0.6{\arcsec}$). 
Note that the mask expansion was not done for sources within $10{\arcsec}$ of
the centre of the galaxy.
Bad and empty pixels were masked by masking pixels with 0 values in the
exposure time maps.
The masks were then manually tweaked to mask scattered light and ghosts from bright stars, or inadequate automatic masking of neighbours.
As the galaxies are all brighter in the red, the position of the
F475W/F475X fit was fixed to the fitted F814W position.
We also adopt the F814W values when reporting results.

The best-fit parameters are the 50\% percentile
from the posterior distributions, with 16\% and 84\% uncertainties.
Model images were created by passing the best-fit parameters to
\texttt{Imfit}'s \texttt{makeimage} executable.
The model image was then subtracted off to create a residual image.
Figure \ref{fig:imfit} shows the F814W \texttt{Imfit} results for 5 UDGs.
In each row, from left to right, is shown the original F814W image, the
residual after subtraction of the best-fit model, the model itself, the
original image after masking bad pixels and pixels belonging to other sources
and finally the masked residual image.
Each cutout is 13 kpc on a side (${\sim}36\arcsec$) at the adopted distance of 75 Mpc to the cluster.
The colour is then measured on the model images using a fixed circular aperture
1~kpc in diameter, with F475X 
transformed to F475W using Equation
\ref{eqn:f475w}. Such an aperture provides a reliable colour, albeit for the more central regions. 
The colour uncertainties are estimated from the posterior distribution of total
magnitudes, computed using Equation 2 in \cite{graham2005}.

\section{Results and Discussion}

\begin{table*}
\caption{
    Structural properties of PIPER LSBs. 
    $M_\mathrm{F814W}$ is the F814W absolute magnitude, adopting $d=75~\mathrm{Mpc}$.
    $\langle\mu\rangle_{\mathrm{e},\mathrm{F814W}}$ and
    $\mu_{0,\mathrm{F814W}}$ are the mean surface brightness within
    $R_\mathrm{e}$ and central surface brightness, respectively.
    The $\mathrm{F475W} - \mathrm{F814W}$ colour is measured inside a 1~kpc diameter aperture on the models.
    $R_\mathrm{e}$ is the model effective radius measured along the
    major-axis, $n$ is the S\'{e}rsic index and $b/a$ is the axis ratio (with
    these parameters measured on the F814W images).
    LSB galaxies that meet the UDG definition are listed at the top, followed
    by those that are UDG-like but either more compact with $R_\mathrm{e} <
    1.5~\mathrm{kpc}$, too bright in F475W surface brightness, or missing the
    information needed to make this determination.
    Galaxies whose IDs are marked with {\textdagger} reside in the NGC~1275
    core fields (R33, W74 and W87).
}
\label{tab:imfit}
\begin{tabular}{lrrccccccc}
\hline
ID & $\alpha$ &
$\delta$ &
$M_\mathrm{F814W}$ &
$\langle\mu\rangle_{\mathrm{e}{,}\mathrm{F814W}}$ &
$\mu_{0{,}\mathrm{F814W}}$ &
$\mathrm{F475W}-\mathrm{F814W}$ &
$R_\mathrm{e}$ &
$n$ &
$b/a$
\\
{} &
(J2000.0) &
(J2000.0) &
(mag) &
($\mathrm{mag}/\Box\arcsec$) &
($\mathrm{mag}/\Box\arcsec$) &
(mag) &
(kpc) &
{} &
{} \\
\hline
R5 & 03:17:34.6 & 41:45:21.6 & $-15.94_{-0.12}^{+0.12}$ & $24.56_{-0.05}^{+0.05}$ & $23.79_{-0.09}^{+0.09}$ & $1.59_{-0.36}^{+0.35}$ & $2.72_{-0.06}^{+0.07}$ & $0.78_{-0.03}^{+0.03}$ & $0.80_{-0.01}^{+0.01}$ \\
R6 & 03:16:58.9 & 41:42:10.1 & $-15.91_{-0.07}^{+0.07}$ & $23.76_{-0.03}^{+0.03}$ & $23.10_{-0.05}^{+0.05}$ & $1.57_{-0.19}^{+0.19}$ & $1.78_{-0.02}^{+0.02}$ & $0.71_{-0.02}^{+0.02}$ & $0.87_{-0.01}^{+0.01}$ \\
R14 & 03:17:06.1 & 41:13:03.2 & $-14.45_{-0.20}^{+0.20}$ & $24.76_{-0.08}^{+0.09}$ & $24.23_{-0.14}^{+0.14}$ & $1.34_{-0.42}^{+0.43}$ & $1.54_{-0.05}^{+0.05}$ & $0.62_{-0.04}^{+0.04}$ & $0.76_{-0.03}^{+0.03}$ \\
R15 & 03:17:03.8 & 41:14:55.0 & $-16.63_{-0.07}^{+0.07}$ & $23.73_{-0.03}^{+0.03}$ & $22.67_{-0.06}^{+0.06}$ & $1.55_{-0.18}^{+0.18}$ & $2.31_{-0.03}^{+0.03}$ & $0.97_{-0.02}^{+0.02}$ & $0.98_{-0.01}^{+0.01}$ \\
R16 & 03:18:36.5 & 41:11:32.1 & $-17.23_{-0.09}^{+0.09}$ & $24.26_{-0.04}^{+0.04}$ & $23.31_{-0.07}^{+0.07}$ & $1.68_{-0.33}^{+0.32}$ & $4.57_{-0.09}^{+0.10}$ & $0.90_{-0.02}^{+0.02}$ & $0.71_{-0.01}^{+0.01}$ \\
R20 & 03:20:24.6 & 41:43:28.6 & $-14.95_{-0.16}^{+0.16}$ & $24.51_{-0.06}^{+0.07}$ & $24.01_{-0.11}^{+0.11}$ & $1.02_{-0.37}^{+0.37}$ & $1.72_{-0.05}^{+0.05}$ & $0.60_{-0.03}^{+0.03}$ & $0.77_{-0.02}^{+0.02}$ \\
R21 & 03:20:29.5 & 41:44:51.0 & $-16.46_{-0.07}^{+0.07}$ & $23.61_{-0.03}^{+0.03}$ & $22.54_{-0.06}^{+0.06}$ & $1.63_{-0.24}^{+0.25}$ & $2.16_{-0.03}^{+0.03}$ & $0.97_{-0.02}^{+0.02}$ & $0.85_{-0.01}^{+0.01}$ \\
R23 & 03:19:51.5 & 41:54:35.6 & $-16.44_{-0.10}^{+0.09}$ & $24.06_{-0.04}^{+0.04}$ & $23.33_{-0.07}^{+0.07}$ & $1.70_{-0.22}^{+0.22}$ & $2.58_{-0.05}^{+0.05}$ & $0.76_{-0.02}^{+0.02}$ & $0.89_{-0.01}^{+0.01}$ \\
R25 & 03:18:26.3 & 41:41:52.2 & $-15.52_{-0.14}^{+0.14}$ & $24.75_{-0.06}^{+0.06}$ & $24.19_{-0.10}^{+0.11}$ & $1.11_{-0.25}^{+0.26}$ & $2.69_{-0.06}^{+0.06}$ & $0.65_{-0.03}^{+0.03}$ & $0.66_{-0.02}^{+0.02}$ \\
R33\textdagger & 03:20:25.8 & 41:31:11.9 & $-16.61_{-0.10}^{+0.10}$ & $23.76_{-0.04}^{+0.04}$ & $22.91_{-0.07}^{+0.07}$ & $1.60_{-0.27}^{+0.28}$ & $2.58_{-0.05}^{+0.06}$ & $0.83_{-0.02}^{+0.02}$ & $0.78_{-0.01}^{+0.01}$ \\
R84 & 03:17:24.9 & 41:44:21.5 & $-16.31_{-0.06}^{+0.06}$ & $23.68_{-0.02}^{+0.02}$ & $22.91_{-0.04}^{+0.05}$ & $1.53_{-0.18}^{+0.18}$ & $1.99_{-0.02}^{+0.02}$ & $0.78_{-0.01}^{+0.01}$ & $0.94_{-0.01}^{+0.01}$ \\
W1 & 03:17:00.4 & 41:19:20.6 & $-15.98_{-0.07}^{+0.07}$ & $23.65_{-0.03}^{+0.03}$ & $22.70_{-0.05}^{+0.05}$ & $1.43_{-0.38}^{+0.38}$ & $1.80_{-0.02}^{+0.02}$ & $0.90_{-0.02}^{+0.02}$ & $0.83_{-0.01}^{+0.01}$ \\
W5 & 03:17:10.9 & 41:34:03.6 & $-15.33_{-0.15}^{+0.15}$ & $24.36_{-0.06}^{+0.06}$ & $23.78_{-0.11}^{+0.11}$ & $1.55_{-0.35}^{+0.35}$ & $1.99_{-0.05}^{+0.05}$ & $0.66_{-0.03}^{+0.03}$ & $0.71_{-0.02}^{+0.02}$ \\
W8 & 03:17:19.6 & 41:34:32.0 & $-13.68_{-0.25}^{+0.25}$ & $25.57_{-0.10}^{+0.10}$ & $25.54_{-0.13}^{+0.11}$ & $1.46_{-0.67}^{+0.66}$ & $1.82_{-0.07}^{+0.07}$ & $0.20_{-0.04}^{+0.05}$ & $0.56_{-0.04}^{+0.04}$ \\
W12 & 03:17:36.7 & 41:23:00.9 & $-15.33_{-0.16}^{+0.16}$ & $23.94_{-0.07}^{+0.07}$ & $22.83_{-0.15}^{+0.14}$ & $1.75_{-0.30}^{+0.32}$ & $1.59_{-0.04}^{+0.05}$ & $0.99_{-0.04}^{+0.05}$ & $0.75_{-0.02}^{+0.02}$ \\
W17 & 03:17:44.1 & 41:21:18.7 & $-15.48_{-0.10}^{+0.10}$ & $23.93_{-0.04}^{+0.04}$ & $23.16_{-0.07}^{+0.07}$ & $1.59_{-0.21}^{+0.21}$ & $1.92_{-0.04}^{+0.04}$ & $0.79_{-0.02}^{+0.02}$ & $0.59_{-0.01}^{+0.01}$ \\
W18 & 03:17:48.4 & 41:18:39.3 & $-15.28_{-0.43}^{+0.44}$ & $25.07_{-0.15}^{+0.16}$ & $24.17_{-0.28}^{+0.29}$ & $1.65_{-0.70}^{+0.70}$ & $3.13_{-0.29}^{+0.32}$ & $0.86_{-0.08}^{+0.08}$ & $0.52_{-0.03}^{+0.04}$ \\
W19 & 03:17:53.1 & 41:19:31.9 & $-15.25_{-0.25}^{+0.25}$ & $24.44_{-0.10}^{+0.10}$ & $23.23_{-0.20}^{+0.20}$ & $1.62_{-0.46}^{+0.46}$ & $1.84_{-0.09}^{+0.10}$ & $1.05_{-0.06}^{+0.06}$ & $0.83_{-0.03}^{+0.03}$ \\
W33 & 03:18:25.9 & 41:41:07.3 & $-15.04_{-0.18}^{+0.17}$ & $24.51_{-0.07}^{+0.07}$ & $23.75_{-0.13}^{+0.13}$ & $1.36_{-0.34}^{+0.36}$ & $1.70_{-0.05}^{+0.06}$ & $0.78_{-0.04}^{+0.04}$ & $0.85_{-0.02}^{+0.02}$ \\
W74\textdagger & 03:19:22.0 & 41:27:23.0 & $-15.71_{-0.10}^{+0.10}$ & $23.93_{-0.04}^{+0.04}$ & $23.34_{-0.07}^{+0.07}$ & $1.64_{-0.29}^{+0.30}$ & $1.70_{-0.03}^{+0.03}$ & $0.67_{-0.02}^{+0.02}$ & $0.93_{-0.02}^{+0.02}$ \\
W79 & 03:19:39.2 & 41:12:05.6 & $-15.47_{-0.12}^{+0.13}$ & $24.36_{-0.05}^{+0.05}$ & $23.42_{-0.10}^{+0.10}$ & $1.46_{-0.27}^{+0.26}$ & $2.22_{-0.05}^{+0.05}$ & $0.89_{-0.03}^{+0.03}$ & $0.65_{-0.02}^{+0.02}$ \\
W88 & 03:19:59.1 & 41:18:32.4 & $-17.42_{-0.13}^{+0.13}$ & $24.30_{-0.04}^{+0.04}$ & $23.01_{-0.09}^{+0.08}$ & $1.49_{-0.32}^{+0.31}$ & $5.44_{-0.17}^{+0.17}$ & $1.10_{-0.02}^{+0.02}$ & $0.61_{-0.01}^{+0.01}$ \\
W89 & 03:20:00.1 & 41:17:05.4 & $-14.81_{-0.15}^{+0.15}$ & $24.70_{-0.06}^{+0.06}$ & $24.17_{-0.11}^{+0.11}$ & $1.20_{-0.44}^{+0.44}$ & $1.83_{-0.05}^{+0.05}$ & $0.63_{-0.03}^{+0.03}$ & $0.71_{-0.02}^{+0.02}$ \\
\hline
R27 & 03:19:43.6 & 41:42:46.8 & $-17.12_{-0.04}^{+0.04}$ & $23.05_{-0.01}^{+0.02}$ & $22.23_{-0.03}^{+0.03}$ & $1.61_{-0.09}^{+0.09}$ & $2.23_{-0.02}^{+0.02}$ & $0.82_{-0.01}^{+0.01}$ & $0.88_{-0.01}^{+0.01}$ \\
R60 & 03:19:36.2 & 41:57:26.2 & $-15.60_{-0.08}^{+0.08}$ & $23.67_{-0.03}^{+0.03}$ & $22.96_{-0.06}^{+0.06}$ & $1.46_{-0.33}^{+0.33}$ & $1.42_{-0.02}^{+0.02}$ & $0.75_{-0.02}^{+0.02}$ & $0.95_{-0.01}^{+0.01}$ \\
R79 & 03:18:21.2 & 41:46:15.3 & -- & -- & -- & -- & -- & -- & -- \\
R89 & 03:20:12.8 & 41:44:57.7 & $-14.03_{-0.11}^{+0.11}$ & $23.47_{-0.04}^{+0.05}$ & $22.84_{-0.09}^{+0.09}$ & $1.05_{-0.24}^{+0.25}$ & $0.64_{-0.01}^{+0.01}$ & $0.69_{-0.03}^{+0.03}$ & $0.91_{-0.02}^{+0.02}$ \\
R116 & 03:17:46.0 & 41:30:11.6 & $-14.55_{-0.13}^{+0.13}$ & $23.85_{-0.05}^{+0.05}$ & $23.31_{-0.09}^{+0.09}$ & $1.46_{-0.37}^{+0.35}$ & $0.96_{-0.02}^{+0.02}$ & $0.63_{-0.03}^{+0.03}$ & $0.92_{-0.02}^{+0.03}$ \\
R117 & 03:18:03.8 & 41:27:08.8 & $-14.47_{-0.13}^{+0.13}$ & $23.41_{-0.05}^{+0.05}$ & $22.78_{-0.10}^{+0.09}$ & $1.42_{-0.35}^{+0.34}$ & $0.91_{-0.02}^{+0.02}$ & $0.69_{-0.03}^{+0.03}$ & $0.64_{-0.02}^{+0.02}$ \\
W2 & 03:17:03.3 & 41:20:28.7 & $-15.05_{-0.74}^{+0.67}$ & $26.06_{-0.28}^{+0.30}$ & $22.95_{-0.64}^{+0.64}$ & -- & $3.70_{-0.51}^{+0.76}$ & $2.09_{-0.18}^{+0.19}$ & $0.76_{-0.04}^{+0.04}$ \\
W4 & 03:17:07.1 & 41:22:52.4 & $-15.55_{-0.14}^{+0.14}$ & $24.00_{-0.06}^{+0.06}$ & $22.57_{-0.12}^{+0.12}$ & $1.48_{-0.63}^{+0.61}$ & $1.67_{-0.05}^{+0.05}$ & $1.18_{-0.03}^{+0.03}$ & $0.88_{-0.02}^{+0.02}$ \\
W6 & 03:17:13.3 & 41:22:07.5 & $-13.64_{-0.21}^{+0.20}$ & $24.15_{-0.08}^{+0.08}$ & $23.70_{-0.14}^{+0.14}$ & $1.60_{-0.61}^{+0.63}$ & $0.83_{-0.03}^{+0.03}$ & $0.57_{-0.04}^{+0.04}$ & $0.71_{-0.03}^{+0.03}$ \\
W7 & 03:17:16.0 & 41:20:11.7 & $-14.99_{-0.08}^{+0.08}$ & $23.69_{-0.03}^{+0.03}$ & $23.21_{-0.05}^{+0.05}$ & $1.49_{-0.24}^{+0.23}$ & $1.17_{-0.02}^{+0.02}$ & $0.59_{-0.02}^{+0.02}$ & $0.81_{-0.01}^{+0.01}$ \\
W13 & 03:17:38.2 & 41:31:56.6 & $-14.71_{-0.19}^{+0.19}$ & $24.32_{-0.08}^{+0.08}$ & $23.20_{-0.16}^{+0.15}$ & $1.41_{-0.46}^{+0.47}$ & $1.31_{-0.04}^{+0.05}$ & $1.00_{-0.05}^{+0.05}$ & $0.89_{-0.03}^{+0.03}$ \\
W14 & 03:17:39.2 & 41:31:03.5 & -- & -- & -- & -- & -- & -- & -- \\
W16 & 03:17:41.8 & 41:24:02.0 & $-13.91_{-0.22}^{+0.23}$ & $24.70_{-0.09}^{+0.09}$ & $24.27_{-0.16}^{+0.15}$ & $1.59_{-0.54}^{+0.52}$ & $1.31_{-0.05}^{+0.05}$ & $0.56_{-0.04}^{+0.05}$ & $0.61_{-0.03}^{+0.03}$ \\
W22 & 03:18:05.4 & 41:27:42.1 & $-15.42_{-0.12}^{+0.12}$ & $24.45_{-0.05}^{+0.05}$ & $23.81_{-0.10}^{+0.10}$ & -- & $2.11_{-0.04}^{+0.05}$ & $0.70_{-0.03}^{+0.03}$ & $0.74_{-0.02}^{+0.02}$ \\
W25 & 03:18:15.5 & 41:28:35.3 & $-14.80_{-0.13}^{+0.13}$ & $23.77_{-0.05}^{+0.06}$ & $23.16_{-0.09}^{+0.10}$ & $1.51_{-0.38}^{+0.36}$ & $1.31_{-0.03}^{+0.03}$ & $0.68_{-0.03}^{+0.03}$ & $0.59_{-0.02}^{+0.02}$ \\
W28 & 03:18:21.7 & 41:45:27.5 & $-13.30_{-0.21}^{+0.21}$ & $24.13_{-0.08}^{+0.08}$ & $23.87_{-0.13}^{+0.13}$ & $1.36_{-0.69}^{+0.71}$ & $0.63_{-0.02}^{+0.02}$ & $0.42_{-0.04}^{+0.04}$ & $0.89_{-0.04}^{+0.04}$ \\
W29 & 03:18:23.3 & 41:45:00.6 & -- & -- & -- & -- & -- & -- & -- \\
W35 & 03:18:28.3 & 41:39:48.5 & $-14.60_{-0.31}^{+0.31}$ & $25.13_{-0.11}^{+0.11}$ & $24.29_{-0.20}^{+0.21}$ & -- & $1.83_{-0.11}^{+0.11}$ & $0.83_{-0.06}^{+0.06}$ & $0.87_{-0.05}^{+0.06}$ \\
W36 & 03:18:29.2 & 41:41:38.9 & -- & -- & -- & -- & -- & -- & -- \\
W40 & 03:18:33.3 & 41:40:55.9 & $-15.00_{-0.13}^{+0.13}$ & $23.93_{-0.05}^{+0.05}$ & $23.38_{-0.09}^{+0.09}$ & $1.36_{-0.29}^{+0.28}$ & $1.39_{-0.03}^{+0.04}$ & $0.64_{-0.03}^{+0.03}$ & $0.72_{-0.02}^{+0.02}$ \\
W41 & 03:18:33.6 & 41:41:58.3 & $-14.97_{-0.31}^{+0.32}$ & $24.50_{-0.13}^{+0.13}$ & $22.05_{-0.30}^{+0.30}$ & -- & $1.57_{-0.10}^{+0.11}$ & $1.74_{-0.09}^{+0.09}$ & $0.94_{-0.03}^{+0.03}$ \\
W56 & 03:18:48.1 & 41:14:02.2 & $-13.83_{-0.19}^{+0.19}$ & $24.71_{-0.08}^{+0.08}$ & $24.39_{-0.12}^{+0.12}$ & $1.52_{-0.68}^{+0.67}$ & $1.15_{-0.03}^{+0.04}$ & $0.48_{-0.03}^{+0.04}$ & $0.74_{-0.03}^{+0.03}$ \\
W59 & 03:18:54.3 & 41:15:28.9 & $-14.71_{-0.09}^{+0.09}$ & $23.65_{-0.04}^{+0.04}$ & $23.07_{-0.07}^{+0.07}$ & $1.31_{-0.29}^{+0.29}$ & $0.97_{-0.02}^{+0.02}$ & $0.66_{-0.02}^{+0.02}$ & $0.88_{-0.02}^{+0.02}$ \\
W80 & 03:19:39.2 & 41:13:43.5 & -- & -- & -- & -- & -- & -- & -- \\
W83 & 03:19:47.4 & 41:44:08.8 & $-13.37_{-0.31}^{+0.31}$ & $24.89_{-0.12}^{+0.13}$ & $24.36_{-0.23}^{+0.22}$ & $1.24_{-0.58}^{+0.59}$ & $1.00_{-0.05}^{+0.05}$ & $0.63_{-0.07}^{+0.07}$ & $0.75_{-0.05}^{+0.05}$ \\
W84 & 03:19:49.7 & 41:43:42.4 & $-13.97_{-0.21}^{+0.20}$ & $23.82_{-0.09}^{+0.09}$ & $22.64_{-0.18}^{+0.19}$ & $1.29_{-0.38}^{+0.39}$ & $0.81_{-0.03}^{+0.03}$ & $1.03_{-0.06}^{+0.06}$ & $0.74_{-0.02}^{+0.03}$ \\
W87\textdagger & 03:19:57.4 & 41:29:31.4 & $-12.98_{-0.16}^{+0.16}$ & $23.67_{-0.06}^{+0.06}$ & $23.48_{-0.09}^{+0.08}$ & $1.32_{-0.45}^{+0.46}$ & $0.44_{-0.01}^{+0.01}$ & $0.37_{-0.02}^{+0.03}$ & $0.88_{-0.03}^{+0.03}$ \\
\hline

\end{tabular}
\end{table*}

\begin{table*}
\caption{
    Globular cluster properties of PIPER LSBs.
    $M_\mathrm{GC}/M_{\star}$ is the ratio of GC system mass to galaxy stellar mass.
    $R_\mathrm{GC}$ is the half-number radius of the GC system.
    $R_\mathrm{GC}/R_\mathrm{e{,}c}$ is the ratio of GC system to host galaxy (circularized) size.
    $\langle\mathrm{F475W} - \mathrm{F814W}\rangle$ is the mean colour of the GCs.
    The last column shows the results of a visual determination of the presence of a stellar nucleus.
    R33, W74 and W87 reside in the NGC~1275 core fields where no GC assessment
    was performed, and are marked with a {\textdagger}.
}
\label{tab:GCs}
\begin{tabular}{lcccrcc}
\hline
ID &
$N_\mathrm{GC}$ &
$M_\mathrm{GC}/M_{\star}$ & 
$R_\mathrm{GC}$ &
$R_\mathrm{GC}/R_\mathrm{e{,}c}$ &
$\langle\mathrm{F475W} - \mathrm{F814W}\rangle$ &
Nucleated
\\
{} &
{} &
{} &
{(kpc)} &
{} &
{(mag)} &
{}
\\
\hline
R5 & $13 \pm 6$ & $1.0\%_{-0.5}^{+0.6}$ & $2.9_{-0.7}^{+1.0}$ & $1.2_{-0.3}^{+0.4}$ & $1.68 \pm 0.07$ & No \\
R6 & $7 \pm 1$ & $0.5\%_{-0.1}^{+0.1}$ & -- & -- & -- & No \\
R14 & $-1 \pm 5$ & $0.0\%_{-0.0}^{+0.0}$ & -- & -- & -- & No \\
R15 & $13 \pm 5$ & $0.5\%_{-0.2}^{+0.2}$ & $2.6_{-0.7}^{+1.0}$ & $1.2_{-0.3}^{+0.5}$ & $1.35 \pm 0.03$ & Yes \\
R16 & $23 \pm 7$ & $0.5\%_{-0.2}^{+0.2}$ & $3.7_{-0.7}^{+0.9}$ & $1.0_{-0.2}^{+0.3}$ & $1.61 \pm 0.04$ & No \\
R20 & $-2 \pm 8$ & $0.0\%_{-0.0}^{+0.0}$ & -- & -- & -- & No \\
R21 & $36 \pm 8$ & $1.7\%_{-0.4}^{+0.5}$ & $3.1_{-0.5}^{+0.6}$ & $1.5_{-0.3}^{+0.3}$ & $1.51 \pm 0.03$ & Yes \\
R23 & $9 \pm 6$ & $0.4\%_{-0.3}^{+0.3}$ & $2.5_{-0.7}^{+1.1}$ & $1.0_{-0.3}^{+0.5}$ & $1.56 \pm 0.05$ & No \\
R25 & $-8 \pm 11$ & $0.0\%_{-0.0}^{+0.0}$ & -- & -- & -- & No \\
R33\textdagger & -- & -- & -- & -- & -- & No \\
R84 & $43 \pm 6$ & $2.2\%_{-0.4}^{+0.5}$ & $2.2_{-0.3}^{+0.4}$ & $1.2_{-0.2}^{+0.2}$ & $1.45 \pm 0.03$ & No \\
W1 & $8 \pm 10$ & $0.6\%_{-0.6}^{+0.8}$ & $3.9_{-1.2}^{+1.9}$ & $2.4_{-0.8}^{+1.2}$ & $1.57 \pm 0.09$ & No \\
W5 & $9 \pm 7$ & $1.2\%_{-0.9}^{+1.2}$ & $3.4_{-0.9}^{+1.4}$ & $2.0_{-0.6}^{+0.9}$ & $1.36 \pm 0.07$ & No \\
W8 & $-0 \pm 7$ & $0.0\%_{-0.0}^{+0.0}$ & -- & -- & -- & No \\
W12 & $2 \pm 7$ & $0.2\%_{-0.2}^{+1.1}$ & -- & -- & -- & No \\
W17 & $7 \pm 6$ & $0.8\%_{-0.7}^{+0.9}$ & $3.8_{-1.1}^{+1.8}$ & $2.5_{-0.8}^{+1.3}$ & $1.72 \pm 0.14$ & No \\
W18 & $-4 \pm 9$ & $0.0\%_{-0.0}^{+0.0}$ & -- & -- & -- & No \\
W19 & $-5 \pm 6$ & $0.0\%_{-0.0}^{+0.0}$ & -- & -- & -- & No \\
W33 & $28 \pm 12$ & $4.7\%_{-2.4}^{+3.1}$ & $2.4_{-0.4}^{+0.5}$ & $1.5_{-0.3}^{+0.4}$ & $1.30 \pm 0.03$ & No \\
W74\textdagger & -- & -- & -- & -- & -- & No \\
W79 & $6 \pm 8$ & $0.6\%_{-0.6}^{+1.0}$ & -- & -- & -- & No \\
W88 & $24 \pm 13$ & $0.4\%_{-0.3}^{+0.3}$ & $3.5_{-0.6}^{+0.8}$ & $0.8_{-0.2}^{+0.2}$ & $1.64 \pm 0.07$ & No \\
W89 & $17 \pm 14$ & $3.4\%_{-2.9}^{+3.8}$ & $4.0_{-0.9}^{+1.2}$ & $2.6_{-0.7}^{+0.9}$ & $1.42 \pm 0.08$ & No \\
\hline
R27 & $52 \pm 8$ & $1.3\%_{-0.2}^{+0.2}$ & $2.4_{-0.3}^{+0.4}$ & $1.1_{-0.2}^{+0.2}$ & $1.44 \pm 0.02$ & No \\
R60 & $20 \pm 6$ & $2.0\%_{-0.7}^{+0.8}$ & $2.6_{-0.5}^{+0.7}$ & $1.9_{-0.4}^{+0.5}$ & $1.37 \pm 0.05$ & Yes \\
R79 & $-4 \pm 9$ & -- & -- & -- & -- & Yes \\
R89 & $-1 \pm 7$ & $0.0\%_{-0.0}^{+0.0}$ & -- & -- & -- & No \\
R116 & $4 \pm 8$ & $1.2\%_{-1.2}^{+2.4}$ & -- & -- & -- & No \\
R117 & $1 \pm 8$ & $0.3\%_{-0.3}^{+2.4}$ & -- & -- & -- & Yes \\
W2 & $1 \pm 10$ & $0.1\%_{-0.1}^{+3.0}$ & -- & -- & -- & Yes \\
W4 & $3 \pm 12$ & $0.3\%_{-0.3}^{+1.5}$ & -- & -- & -- & No \\
W6 & $-2 \pm 10$ & $0.0\%_{-0.0}^{+0.0}$ & -- & -- & -- & Yes \\
W7 & $0 \pm 10$ & $0.0\%_{-0.0}^{+1.8}$ & -- & -- & -- & No \\
W13 & $1 \pm 7$ & $0.2\%_{-0.2}^{+2.0}$ & -- & -- & -- & Yes \\
W14 & $-1 \pm 7$ & -- & -- & -- & -- & No \\
W16 & $-2 \pm 7$ & $0.0\%_{-0.0}^{+0.0}$ & -- & -- & -- & No \\
W22 & $-2 \pm 8$ & $0.0\%_{-0.0}^{+0.0}$ & -- & -- & -- & No \\
W25 & $4 \pm 6$ & $0.8\%_{-0.8}^{+1.6}$ & -- & -- & -- & No \\
W28 & $2 \pm 9$ & $1.9\%_{-1.9}^{+9.3}$ & -- & -- & -- & No \\
W29 & $9 \pm 9$ & -- & $4.4_{-1.2}^{+1.8}$ & -- & $1.83 \pm 0.13$ & Yes \\
W35 & $2 \pm 11$ & $0.6\%_{-0.6}^{+3.8}$ & -- & -- & -- & No \\
W36 & $-6 \pm 12$ & -- & -- & -- & -- & No \\
W40 & $13 \pm 10$ & $2.2\%_{-1.8}^{+2.3}$ & $3.9_{-1.0}^{+1.4}$ & $3.3_{-0.9}^{+1.3}$ & $1.94 \pm 0.12$ & No \\
W41 & $-1 \pm 12$ & $0.0\%_{-0.0}^{+0.0}$ & -- & -- & -- & No \\
W56 & $4 \pm 8$ & $1.9\%_{-1.9}^{+5.2}$ & -- & -- & -- & No \\
W59 & $0 \pm 9$ & $0.1\%_{-0.1}^{+2.3}$ & -- & -- & -- & Yes \\
W80 & $7 \pm 8$ & -- & $4.3_{-1.3}^{+2.0}$ & -- & $1.29 \pm 0.10$ & No \\
W83 & $6 \pm 7$ & $4.5\%_{-4.5}^{+8.7}$ & -- & -- & -- & No \\
W84 & $1 \pm 7$ & $0.3\%_{-0.3}^{+3.8}$ & -- & -- & -- & No \\
W87\textdagger & -- & -- & -- & -- & -- & No \\
\hline

\end{tabular}
\end{table*}

\subsection{Galaxy Structural Properties}

Table \ref{tab:imfit} lists the structural properties of the 50 sample
galaxies, including coordinates, total magnitudes, surface brightnesses (mean
and central), colours within 1 kpc, and S\'{e}rsic profile parameters (i.e.\
half-light radii, S\'{e}rsic index and axis ratios $b/a$).
All parameters were measured in F814W, with adopted values being the 50\%
percentiles from the \texttt{Imfit} posterior distributions, with 16\% and 84\%
percentile uncertainties.
The MCMC mode of \texttt{Imfit} was chosen as it provides slightly larger and
more realistic uncertainties than $\chi^2$ minimization.
While the uncertainties may still be underestimated, we note that the more
important quantity of interest is the ratio of the size of the GC system
($R_\mathrm{GC}$) to the galaxy, for which the uncertainty will be dominated
by $R_\mathrm{GC}$ (see \S \ref{sec:R_GC}).

Five galaxies (R79, W14, W29, W36 and W80) could not be modelled successfully
with \texttt{Imfit}.
This was due to bright stars or strong artifacts that resulted in too little
galaxy light after masking, or the galaxy was simply too faint to model.
An additional four galaxies were not fitted in the F475W filter.
Thus structural {\it and} colour information is only available for 41 of the
50 galaxies in the sample.

\begin{figure}
    \includegraphics[width=0.485\textwidth]{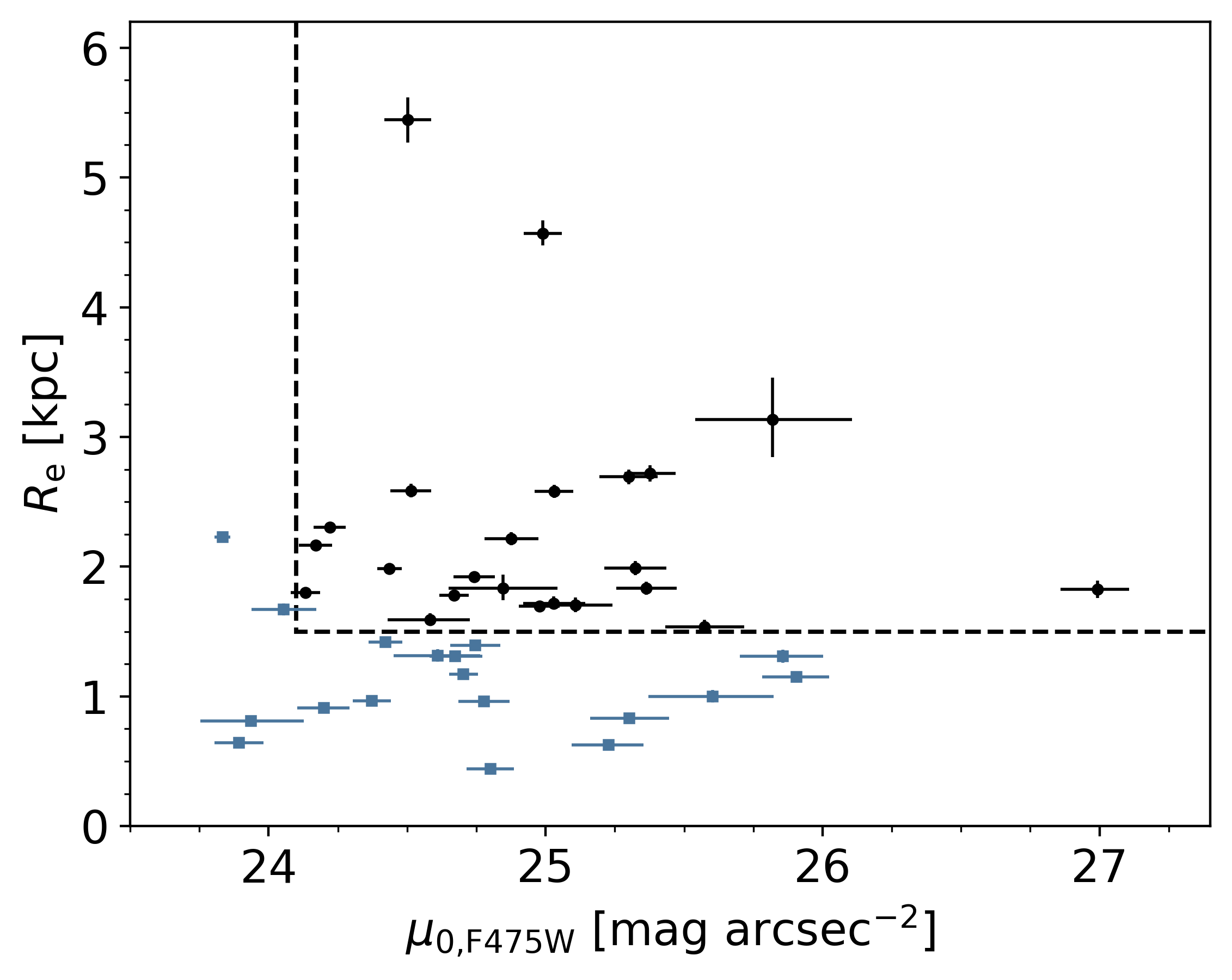}
	\caption{
    Half-light radii and central surface brightnesses for 41 Perseus cluster
    galaxies with both structural information and colours.
    $R_\mathrm{e}$ is the half-light radius along the major-axis from the
    F814W \texttt{Imfit} model, and $\mu_{0{,}\mathrm{F475W}}$ is the central
    surface brightness in F475W, roughly equivalent to SDSS $g$.
    The dashed lines show the canonical UDG criteria of
    $R_\mathrm{e} > 1.5~\mathrm{kpc}$ and $\mu_{0{,}\mathrm{F475W}} >
    24.1~\mathrm{mag}~\mathrm{arcsec}^{-2}$.
    The 23 objects which pass the UDG criteria are plotted as black circles, while blue
    squares show the 18 that do not. 
    \label{fig:size-SB}
	}
\end{figure}

\cite{vandokkum2015} originally defined UDGs as those with $R_\mathrm{e}
\gtrsim 1.5~\mathrm{kpc}$ and $\mu_{0{,}g} \gtrsim
24~\mathrm{AB}~\mathrm{mag}~\mathrm{arcsec}^{-2}$, where $R_\mathrm{e}$ is the half-light 
radius measured along the major axis and $\mu_{g,0}$ is the central surface
brightness in the $g$-band in AB magnitudes.
In Figure \ref{fig:size-SB}, we plot the F814W half-light radius along the
major-axis and the central surface brightness in F475W, which is similar to
SDSS $g$-band.
The F475W central surface brightness is estimated from the F814W central surface
brightness and the model colour with $\mu_{0{,}\mathrm{F475W}} =
\mu_{0{,}\mathrm{F814W}} + (\mathrm{F475W} - \mathrm{F814W})$.
The dashed line shows the UDG selection criteria.
Here we are using Vega magnitudes instead of AB, which amounts to a difference
of $\Delta\mathrm{F475W}_{\mathrm{AB} - \mathrm{VEGA}} = -0.1.$
So in Vega magnitudes, the bright surface brightness limit is
$\mu_{\mathrm{F475W}{,}0} \approx 24.1~\mathrm{mag}~\mathrm{arcsec}^{-2}$.
Of the 50 LSB galaxies listed in Table \ref{tab:imfit}, we plot the 41 that have {\it both} measured
structural parameters and a colour.
Black circles denote the 23 LSB galaxies which strictly pass the UDG criteria, whereas blue squares include 2 UDG-like objects (W4 and R27) that are somewhat brighter and 16 that are 
more compact than UDGs. 
These symbol definitions will be preserved in subsequent figures.
Galaxies that do not meet the strict UDG definition 
(i.e.\ those with $R_\mathrm{e} < 1.5~\mathrm{kpc}$ or $\mu_{0{,}\mathrm{F475W}} < 24.1~\mathrm{mag}~\mathrm{arcsec}^{-2}$, or missing the information to make the determination) are listed at the end of the table.

Since the UDG definition is rather arbitrary, it is perhaps no surprise that it does not identify a distinct population of galaxies but rather a continuous distribution in both size and surface brightness. 
Some of the UDG-like galaxies, outside of the UDG selection box, in 
Figure \ref{fig:size-SB} could be referred to as NUDGEs \citep{forbes2024} since they nudge up against the standard UDG criteria.
Other galaxies in our sample are in the size--luminosity regime of classical dwarf spheroidals.
Below we will refer to the non-UDGs simply as `dwarfs'. All galaxies in our Perseus sample can be considered as LSB galaxies. 

Galaxy colours are measured on the model images within a 1 kpc diameter aperture. Our sample has an average $\mathrm{F475W} - \mathrm{F814W}$ colour of ${\sim}1.5$, indicative of an old passive stellar population, though a small number of galaxies have bluer colours, e.g.\ R20 and R89 with $\mathrm{F475W} - \mathrm{F814W} \sim 1.0$. 
For a more detailed discussion of the stellar populations of Perseus UDGs from spectroscopy, we refer the reader to \cite{2023MNRAS.526.4735F}. Four of the five UDGs studied were found to be relatively old, metal-poor and alpha-element enhanced. 

Galaxy axis ratios hold potentially important clues to the intrinsic
distribution of galaxy shapes. In \cite{pfeffer2024} we investigate how GC
richness varies with $b/a$ for this sample of galaxies, as well as for UDGs in
other environments.

\subsection{GC System Numbers}

\begin{figure}
    \includegraphics[width=0.485\textwidth]{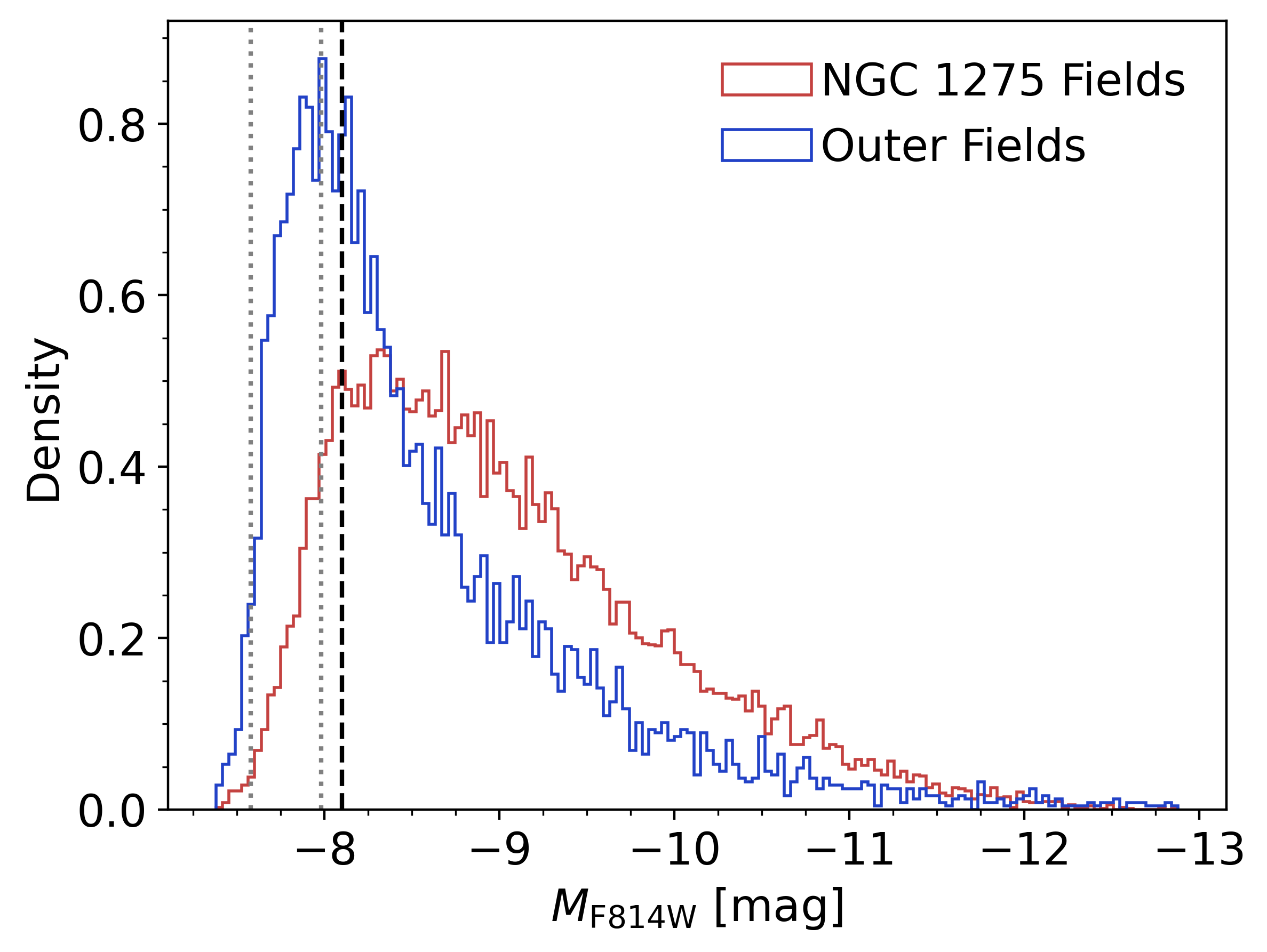}
	\caption{
    Normalized F814W luminosity functions of all GC candidates.
    The red histogram shows the 
    GC candidates from the five NGC~1275 field pairs while the blue histogram are those from the outer fields.
    The dashed black line is the predicted turnover magnitude at $M_{\mathrm{F814W}} \approx
    -8.1$ \citep{2007ApJ...670.1074M}.
    The GC catalog is reasonably complete to the turnover magnitude as shown
    by the dotted grey lines which are the 50\% completeness limits of the
    most and least extincted fields.
    \label{fig:gclf}
	}
\end{figure}

Our GC catalog is reasonably 
complete in magnitude to the expected turnover of
the GCLF of $M_\mathrm{F814W} \approx -8.1$ \citep{2007ApJ...670.1074M}. 
Figure \ref{fig:gclf} shows the magnitude distribution of all GC candidates
from all \textit{HST} pointings, split between the core NGC~1275 fields and the outer fields. The black dashed line shows the predicted
turnover magnitude.
The two dotted grey lines are the 50\% completeness limits of the least and
most extincted fields from the entire survey.
We are thus mostly complete in magnitude  at the turnover magnitude (i.e.,
completeness is between 71\% and 97\%, with a mean completeness of 87\%).
The apparent brighter turnover of the NGC~1275 fields compared to the outer
fields is expected given that the GC host galaxies tend to be giants in the
core and dwarfs in the outer regions \citep{harris2013}.
However, extinction variation between fields also contributes.

Figure \ref{fig:GC_montage} shows $12 \times 12~\mathrm{kpc}^2$ ($33\arcsec \times 33\arcsec$) F814W cutouts for all 50 sample galaxies.
GC candidates are marked in red.
The ID is shown in the upper right of each panel and the total number of GCs,
including both the background and magnitude incompleteness corrections, is shown in the
bottom right.
Background-corrected values less than zero are shown as zero.
The ID is bolded if the galaxy meets the strict definition of a UDG.
Galaxies with IDs marked with {\textdagger} reside in the NGC~1275 core
fields (R33, W74 and W87).

The cutouts reveal a variety of galaxy morphologies and GC system spatial
distributions. For example, R25 is very faint but appears distorted, R89 has
an S-shape which may indicate an interaction or it is perhaps a background
spiral galaxy, as might be W84. W4 may have a bar. Both R21 and R84 have many
GC candidates but the GC system of R84 appears to be more compact spatially.
W6 appears to have a central nucleus but no corresponding GC system. Galaxies
R79, W14, W29, W36 and W80 are all quite faint and a galaxy model could not be
obtained.

\begin{figure*}
    \begin{center}
	\includegraphics[width=0.99\textwidth]{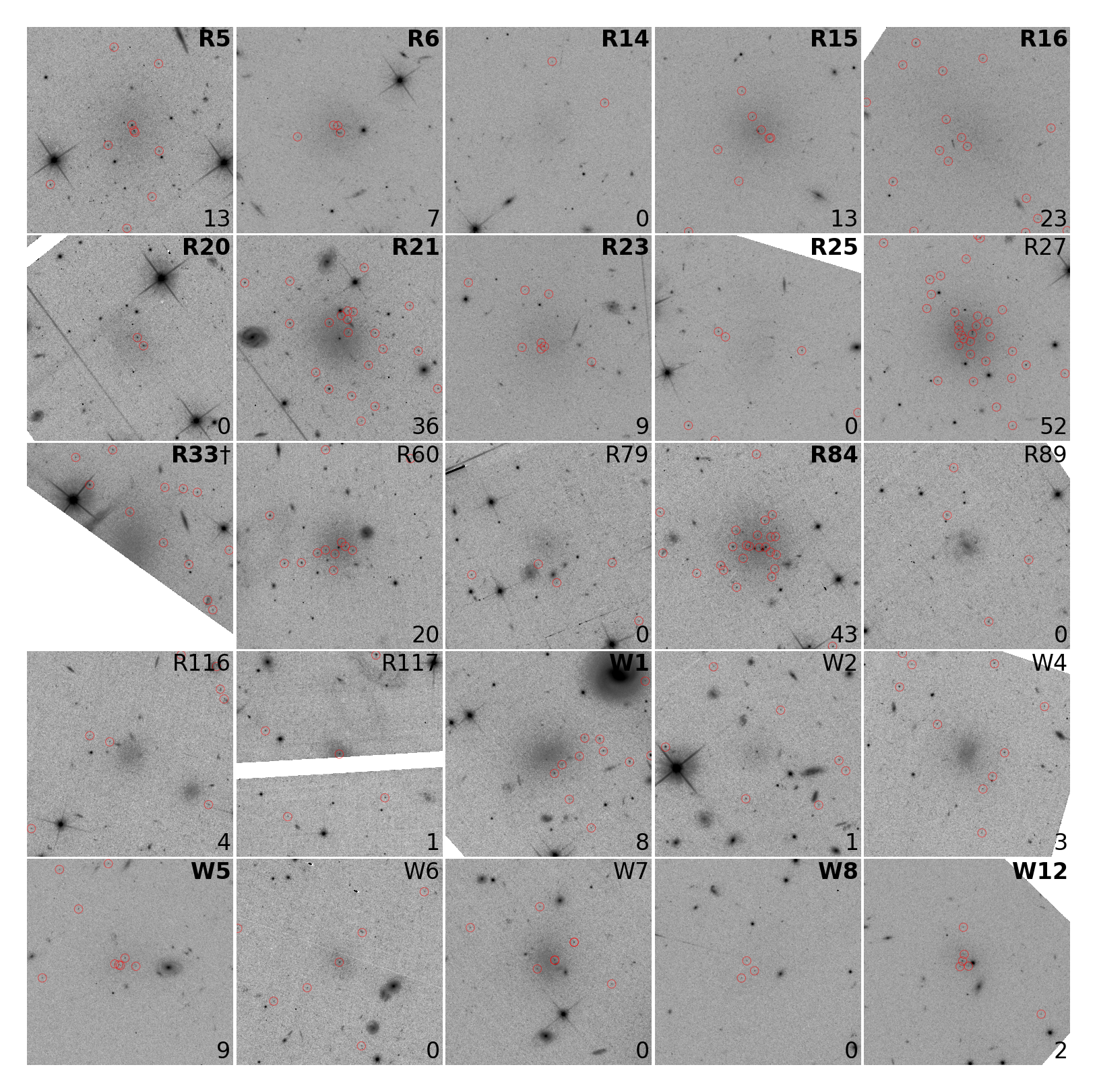}
	\caption{
    F814W postage stamps of all 50 LSB galaxies and their GC candidates.
    The cutouts are $12 \times 12~\mathrm{kpc}^2$ ($33\arcsec \times 33\arcsec$) at the adopted distance to the cluster.
    GC candidates are marked in red.
    Bold IDs denote galaxies that satisfy the strict UDG definition.
    The total number of GCs, after background and magnitude corrections, is
    shown in the bottom right.
    No GC number is shown for the three galaxies in the Perseus core fields near NGC~1275 
    (R33, W74 and W87)
    as the more detailed background modelling required for a GC assessment is
    saved for future work; the IDs for these galaxies are marked with {\textdagger}.
    \label{fig:GC_montage}
	}
    \end{center}
\end{figure*}
\begin{figure*}
    \begin{center}
	\includegraphics[width=0.99\textwidth]{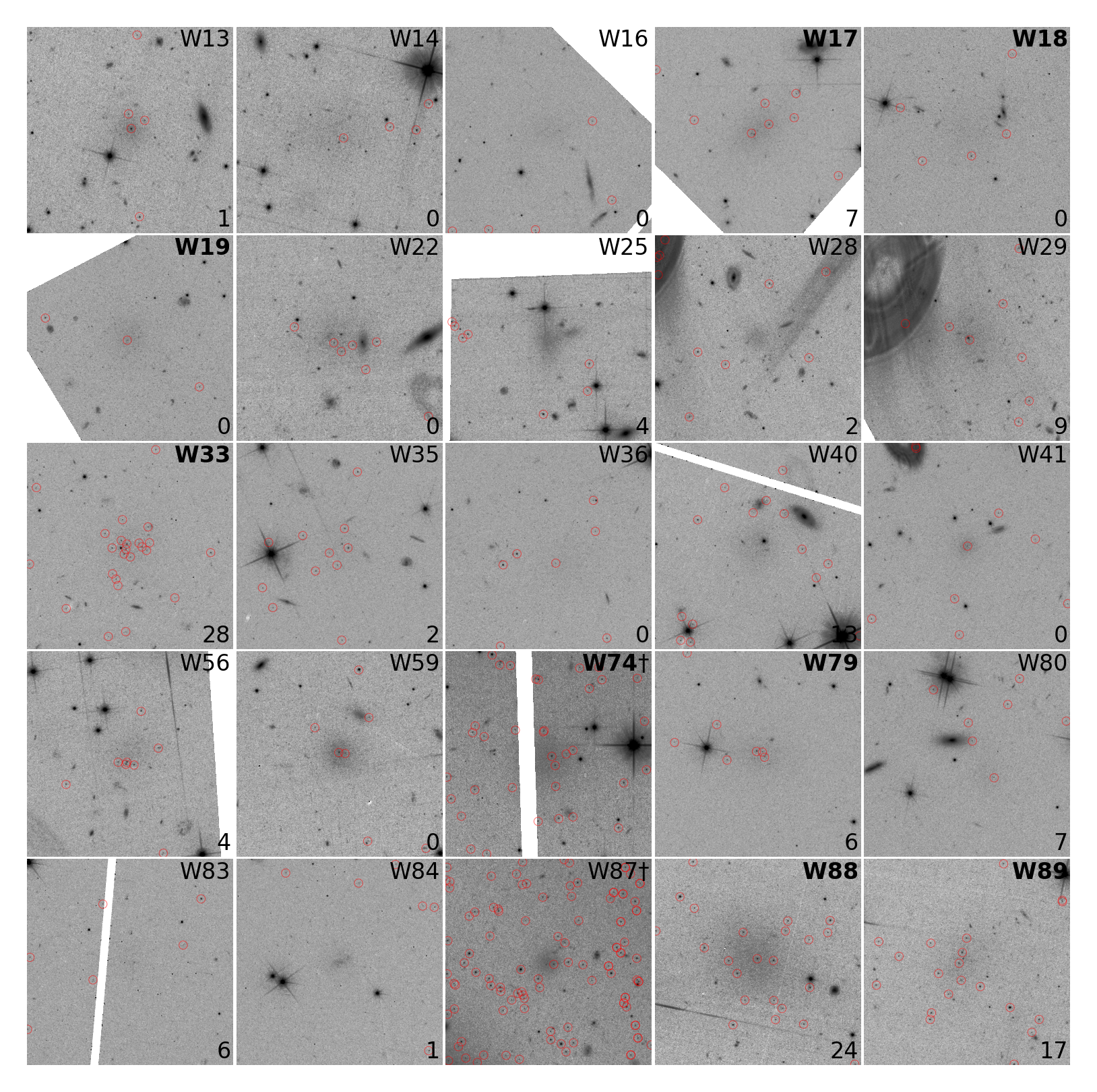}
    \contcaption{}
    \end{center}
\end{figure*}

A GC count for each galaxy was made by first counting the number of GC candidates
in a fixed 15~kpc diameter circular aperture.
Based on the GC system sizes (discussed later), such an aperture is expected
to contain at least 90\% of a galaxy's GCs, and the use of a large fixed-size
aperture decouples the measure of GC numbers from the structural parameters of
the galaxies\footnote{We repeated this procedure using an aperture three times
the host galaxy half-light radius instead and found similar total GC counts
for all galaxies, with the values agreeing within their uncertainties.}.
We then make a correction for background contamination. 
The background density, and standard deviation, are
estimated by counting sources that pass our GC cut in 500 similarly sized
apertures placed randomly around each galaxy's respective image (but outside each
galaxy's GC detection aperture).
The expected number of background contaminants in the galaxy GC aperture, along
with 1~$\sigma$ upper and lower numbers, are computed.
The total number of GCs is then twice the background-corrected number of GCs,
since our GCs only include those brighter than the GCLF turnover magnitude.
We assume a symmetric universal GCLF that has a turnover magnitude of
$M_{\mathrm{TO{,}F814W}} = -8.1$, noting that the GCLF of UDGs is still
somewhat uncertain \citep[e.g.][]{shen2021,janssens2022,2022MNRAS.511.4633S},
and that only NGC~5846\_UDG1 has a near-complete GCLF measured so far
\citep{danieli2022}.
The total GC number uncertainties are computed using the 1 sigma limits of
background contaminants, where for simplicity we assume the uncertainty in the
background dominates the error budget.
Table \ref{tab:GCs} lists the total number of GCs for the 42 galaxies that we were able to measure.
Due to the complex and overwhelming density of GCs near NGC~1275, this
analysis was not done for three galaxies in the NGC~1275 fields: R33, W74 and
W87.
These galaxies do not have $N_\mathrm{GC}$ values listed in Table \ref{tab:GCs} nor shown in Figure \ref{fig:GC_montage}. 
The more detailed background modelling necessary will be left for future work.

The mass of a GC system can be estimated using the total number of GCs and the
typical GC mass, estimated using the GCLF turnover luminosity and a mass-to-light
ratio.
Using the FSPS code \citep{conroy2009, conroy2010}, adopting a simple stellar population (SSP), \cite{chabrier2003} initial mass function, 
$\log(Z/Z_\odot) = -1.5$ and an age of 8~Gyr we find a mass-to-light ratio of $M_\star/L_\mathrm{F814W} =
1.3$. For $M_{\mathrm{TO{,}F814W}} = -8.1$, the typical GC mass is then
${\sim}1 \times 10^5~M_\odot$. Using this mass-to-light ratio for both the GC system and host galaxy, we calculate the ratio of the GC system mass to galaxy stellar mass and include it as a percentage in Table \ref{tab:GCs}. 

\begin{figure}
    \includegraphics[width=0.485\textwidth]{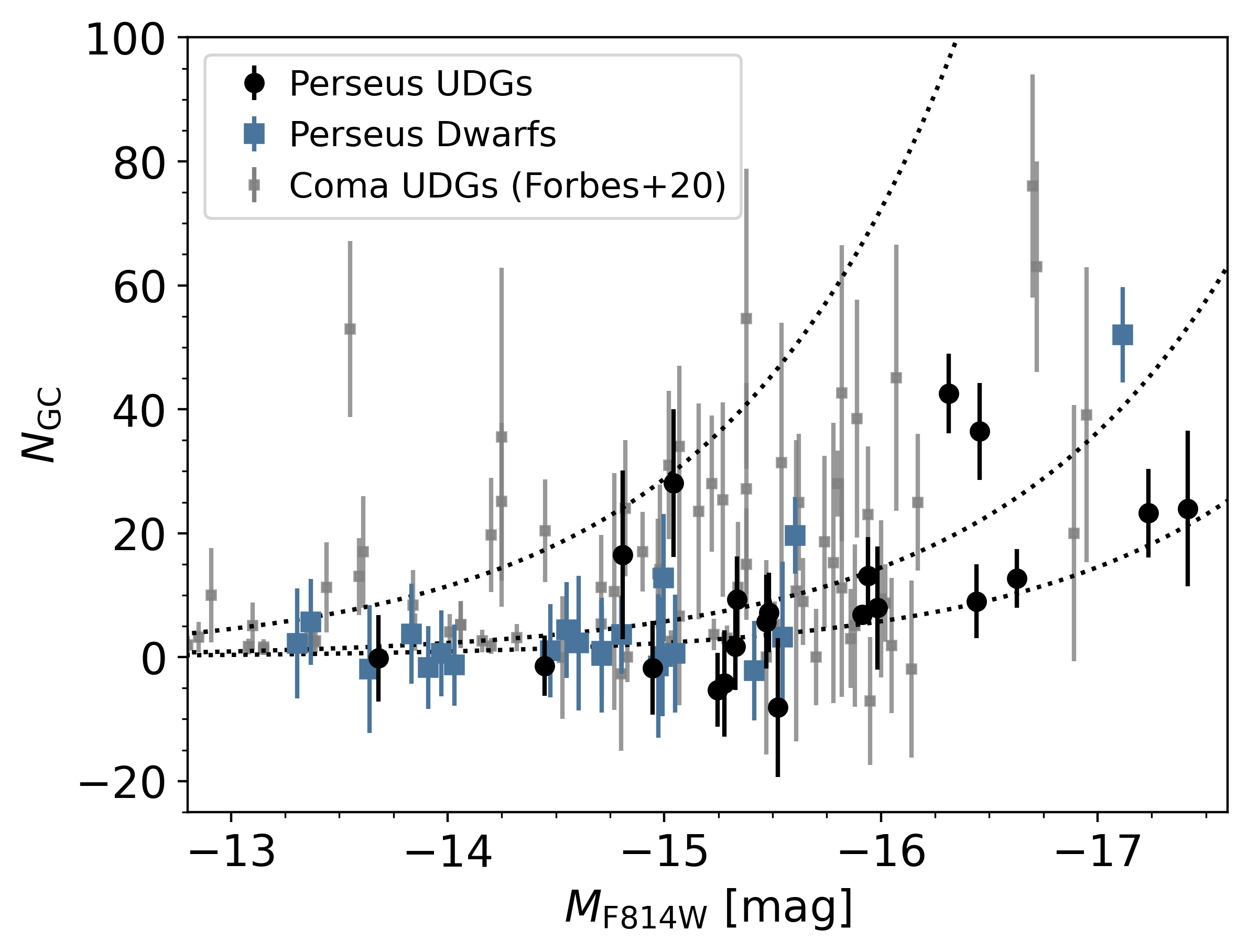}
    \caption{
    Total number of GCs ($N_\mathrm{GC}$) as a function of galaxy
    absolute magnitude.
    Black circles are galaxies which satisfy the UDG
    criteria, blue squares are other LSB dwarfs. The dotted curves show GC
    system mass to stellar mass fractions assuming a mass-to-light ratio
    of $M_\star/L_\mathrm{F814W} = 1.3$ for both. From bottom to top, the
    curves show mass fractions of 0.4\%, 1\% and 5\%. For comparison, we
    also show $N_\mathrm{GC}$ for Coma UDGs from \citealt{forbes2020} in
    grey. The Perseus cluster reveals some UDGs with rich GC systems ($N_\mathrm{GC} > 20$). 
    \label{fig:Ngc_magabs}
    }
\end{figure}

Figure \ref{fig:Ngc_magabs} shows the number of GCs as a function of galaxy F814W luminosity, and directly compares them to UDGs in the Coma cluster. 
As before, black circles are Perseus galaxies from our sample that meet the
UDG criteria, and blue squares are UDG-like galaxies that we designate as dwarfs. 
GC counts for Coma UDGs (grey symbols) are from \cite{forbes2020}, who combined literature values from \cite{vandokkum2017},
\cite{lim2018} and \cite{amorisco2018}.
The Coma cluster UDGs have been converted from $V$-band
luminosities to F814W assuming $V - I = 1$.
The dotted lines show the GC system mass 
to host galaxy stellar mass with both assuming 
$M_\star/L_\mathrm{F814W} = 1.3$.
From bottom to top are plotted GC mass fractions of 0.4\%, 1\% and 5\%.

The plot shows that a few Perseus UDGs are GC rich, hosting many tens of GCs, as is found in the Coma cluster. 
Given the general lack of rich GC systems in other clusters (Fornax, Virgo, Hydra~I) this may suggest that only the most massive ($M_{200} \sim 10^{15}~M_{\odot}$) clusters host GC rich UDGs. 
We note R21 with $36 \pm 8$ and R84 with $43 \pm 6$ GCs. The richest GC system
belongs to R27, which is slightly brighter than the standard UDG definition,
with $52 \pm 8$ GCs. 
The GC system mass fractions for Perseus cluster LSB galaxies are generally consistent with a 0.4\% mass fraction, with the highest being the UDG W33 with a 4.7\% mass fraction. Coma cluster LSB dwarfs reveal a similar range in GC to galaxy mass fractions up to a maximum of 5\% (with a couple of exceptions). 
These ratios can be compared to the Milky Way halo which has a ratio of 1--2\%
\citep{harris2017, deason2019}.
Although our Perseus UDG sample and the Coma UDGs reveal similar GC to galaxy
mass fractions (and possibly higher on average for Coma), the Perseus galaxies
are on average more luminous (higher mass) than the Coma UDG sample from
\cite{forbes2020} and there may be other selection effects that should be
considered.
For example, a weak correlation for GC-rich UDGs to lie at smaller
clustercentric distances is expected \citep{buzzo2024}.
However, this is hard to probe with our dataset as all our UDGs lie at
relatively small projected clustercentric distances, with the most distant
galaxy at ${\sim}750~\mathrm{kpc}$ or ${\sim}0.4 \times R_{200}$.
Moreover, many bright galaxies in the core of Perseus are arranged in a
near-linear feature to the southwest of NGC 1275, and it is possible that the
distance to this filament is the driver of any local environmental effects.

If we separate our Perseus sample into UDGs and dwarfs, we find that the mean
number of GCs hosted by UDGs is $11 \pm 3$ whereas the dwarfs host an average
of only $4 \pm 2$ GCs.
However, in terms of GC system mass to galaxy stellar mass, this difference vanishes with both
the UDGs and dwarfs hosting a similar mean mass fraction. 
Re-examining Figure \ref{fig:size-SB} provides a possible explanation: due to
the degeneracy between effective radius, luminosity and
S\'{e}rsic index, a size cut (i.e.\ the UDG $R_\mathrm{e} > 1.5~\mathrm{kpc}$
definition) in a population of galaxies with similar surface brightnesses and
S\'{e}rsic indices ($n \sim 1$) is equivalent to a luminosity cut, and to a stellar
mass cut assuming similar mass-to-light ratios.
This is reflected in Figure \ref{fig:Ngc_magabs} where most Perseus galaxies brighter
than $M_\mathrm{F814W} = -15$ are UDGs, whereas those fainter are dwarfs.

\subsection{GC System Sizes}\label{sec:R_GC}

\begin{figure*}
    \begin{center}
	\includegraphics[height=1.00\textwidth]{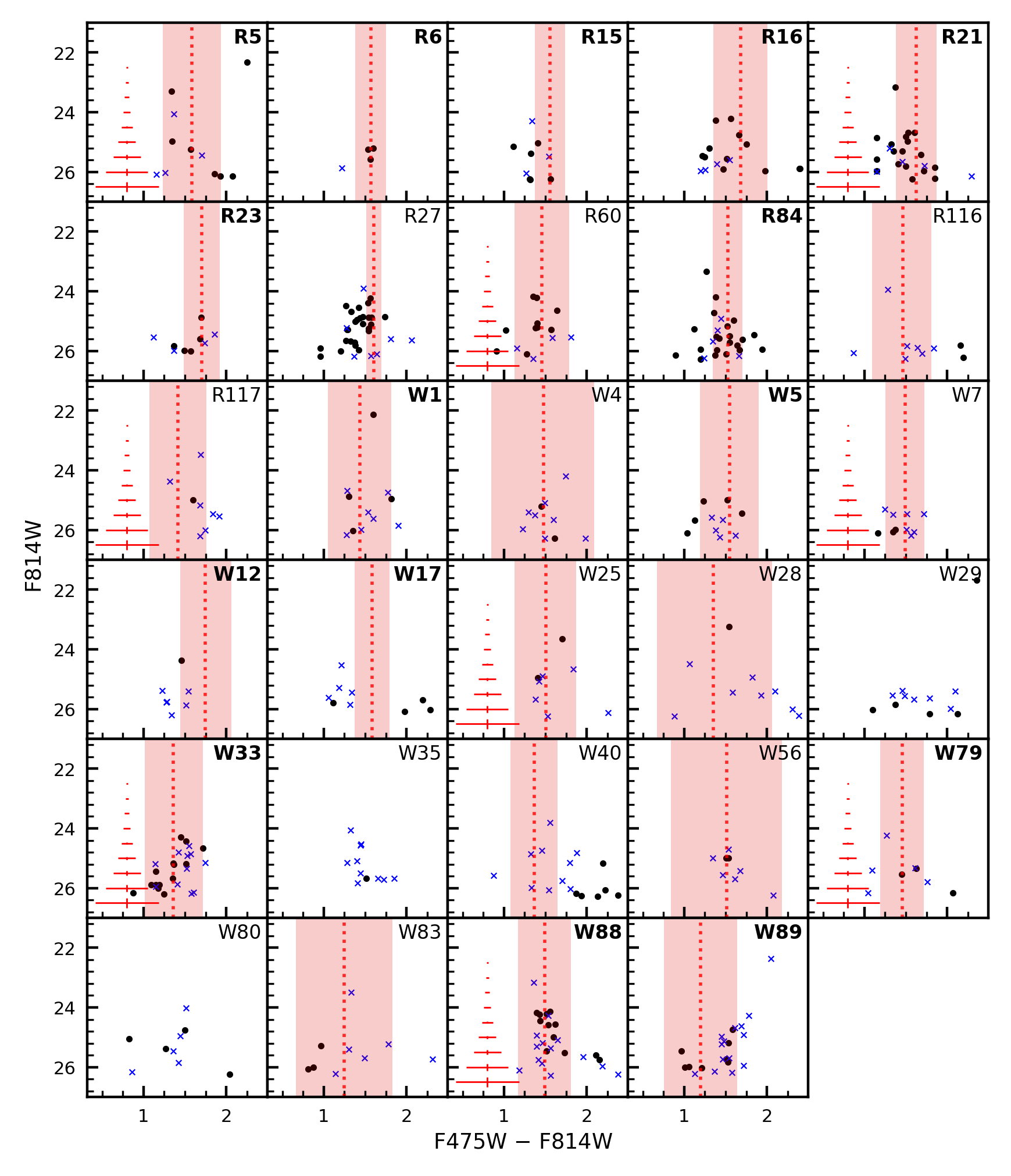}
	\caption{
    Colour--magnitude diagrams of point sources around target galaxies with GC counts above the expected background contribution.
    A statistical background correction was made, removing GC candidates nearest in colour and magnitude (see text for details).
    GC candidates that survived the correction are shown in black and those
    that were removed are the blue crosses.
    Bold IDs denote galaxies that satisfy the strict UDG definition.
    LSBs in the core of the Perseus cluster are also not shown.
    The host galaxy colour is shown as the dotted red vertical line and the shaded band is the colour uncertainty; it is absent if the \texttt{Imfit} result is unreliable.
    Characteristic error bars derived from the artificial star tests are shown on the right side of a subset of panels.
    Many of the GC systems have colours similar to the colour of the host galaxy.
    \label{fig:cmd_montage}
	}
    \end{center}
\end{figure*}

In order to measure the sizes of GC systems we employ 
an 
alternative process for removing background contaminants.
We start with all the GC candidates within a radius $r = 7.5~\mathrm{kpc}$ of each galaxy. 
For each galaxy, the number of expected background contaminants was estimated
in the same way as described above.
A background sample of sources that meet the GC criteria is then randomly
drawn from elsewhere in each galaxy's image.
For each background source, we then removed the galaxy GC candidate nearest in
Euclidean distance on the colour-magnitude diagram.
The result of this process is shown in Figure \ref{fig:cmd_montage}.
Each CMD shows all the GC candidates within $r < 7.5~\mathrm{kpc}$.
Blue $\times$ symbols are candidates that were removed as background objects, and black points
are those that survived this process\footnote{Figure \ref{fig:cmd_montage}
provides a sanity check on the $N_\mathrm{GC}$ values in Table \ref{tab:GCs}.
Twice the number of black points (as only the bright half of the GCLF is
selected and plotted) agree with the total numbers for each galaxy.}.
For comparison, the red dotted line shows the mean colour of the host galaxy, though it is
absent in the cases where \texttt{Imfit} did not converge in at least one of
the bands.

A S\'{e}rsic model was then fitted to the one dimensional radial distribution of
GC candidates after background removal using the \texttt{emcee} package
\citep{foreman-mackey2013}.
The only free parameter is the `effective' radius---which for GCs we will refer to as the
half-number radius $R_\mathrm{GC}$.
The ellipticity of the GC system is fixed to zero.
The S\'{e}rsic index $n$ was initially allowed to vary.
For the richest six GC systems with $N_\mathrm{GC} > 20$ (i.e.\ at least 10
candidates) we found a mean $n$ of $1.1 \pm 0.1$, and subsequently we fixed it
to $n = 1$.
The best-fit $R_\mathrm{GC}$ is the 50\% percentile from the posterior
distribution, with 16\% and 84\% uncertainties.
The 16\%, 50\% and 84\% percentiles from the $R_\mathrm{GC}$ posterior
distribution is recorded.

The background removal and \texttt{emcee} fitting was then repeated 25 times
for each galaxy (Figure \ref{fig:cmd_montage} is showing one realization).
We adopt the mean 50\% $R_\mathrm{GC}$ percentile from the 25 fits as the final
$R_\mathrm{GC}$ value, with the mean 16\% and 84\% percentiles as uncertainties.
Table \ref{tab:GCs} lists the resulting half-number radii ($R_\mathrm{GC}$) in kpc 
and the ratio of GC system size to galaxy half-light radius
($R_\mathrm{GC}/R_\mathrm{e{,}c}$).
The $R_\mathrm{GC}$ posterior distributions will not all overlap when there
are fewer than 4 GC candidates (after background removal), and so
$R_\mathrm{GC}$ and $R_\mathrm{GC}/R_\mathrm{e{,}c}$ values are not reported
in such cases.

As few as 4 GC candidates will still provide a reliable $R_\mathrm{GC}$
estimate.
This was tested on R27 (the galaxy with the richest GC system in our sample)
by re-performing the \texttt{emcee} fit 25 times with a randomly selected
subset of 4 of its 26 candidates.
A majority of the fits agree within uncertainties with the
$2.4^{+0.4}_{-0.3}~\mathrm{kpc}$ estimate, and the mean percentiles overlap.

As noted in the Introduction, the number of GCs around UDGs in the Coma cluster has been the subject of recent debate with much lower GC numbers claimed by \cite{2022MNRAS.511.4633S} for six Coma UDGs.
A key  factor in the different GC numbers found with previous studies is the smaller half-number radii for GC systems found by \cite{2022MNRAS.511.4633S}. For example, 
for DF44 and DFX1 \cite{vandokkum2017} found $R_\mathrm{GC}/ R_\mathrm{e} = 1.5$ whereas \cite{2022MNRAS.511.4633S} found 
$R_\mathrm{GC} / R_\mathrm{e} \sim0.8$ for both. 

\begin{figure*}
    \includegraphics[width=1.0\textwidth]{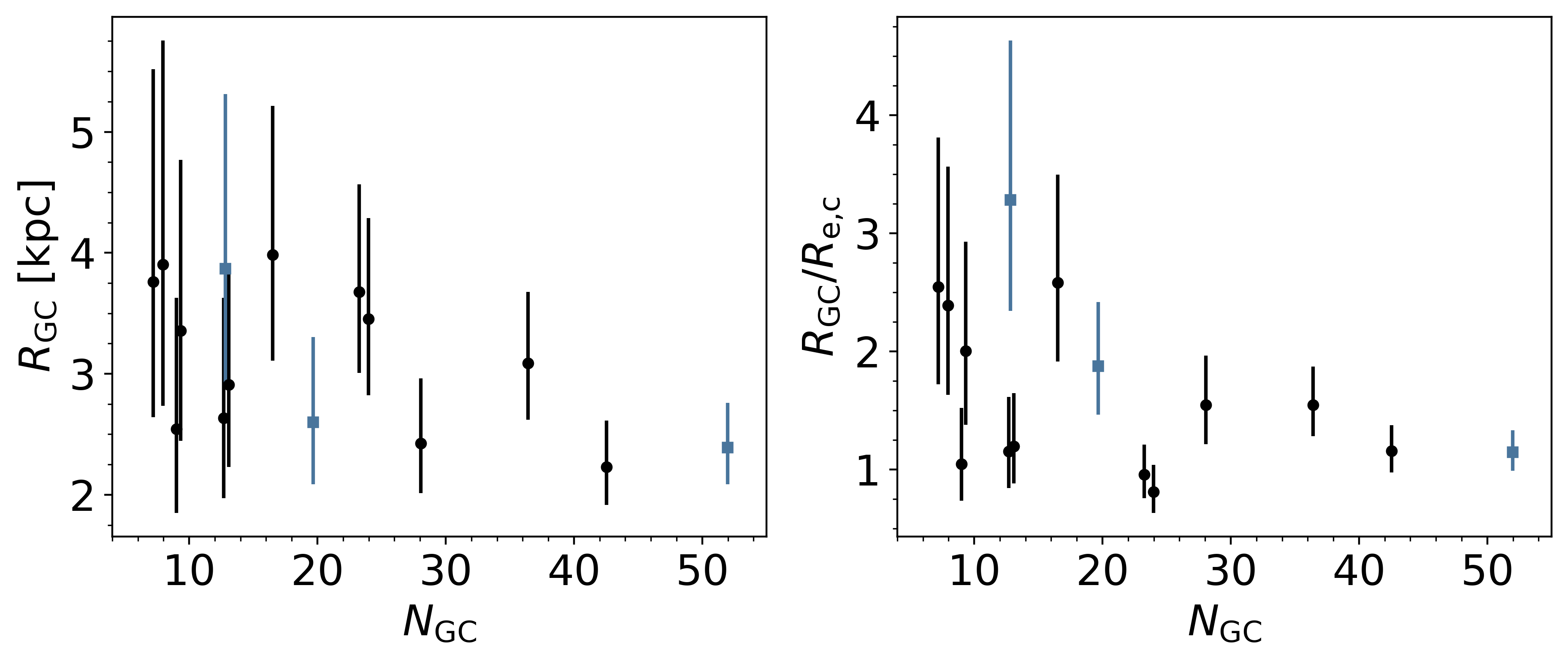}
    \caption{
    \textit{Left}: GC system half-number radius as a function of number of
    GCs.
    \textit{Right}: Ratio of GC half number radius to circularized galaxy
    half light radius plotted against number of GCs.
    Galaxies with the richest GC systems ($N_\mathrm{GC} > 20$) have    
    $R_\mathrm{GC}/R_\mathrm{e}$ ratios of ${\sim}1.2$.
    In both panels only galaxies with at least three GC candidates are included.
    Black circles are galaxies which satisfy the UDG criteria, blue squares are dwarfs. 
    \label{fig:RgcRe}
    }
\end{figure*}

In Figure 
\ref{fig:RgcRe} 
we show both $R_\mathrm{GC}$ and $R_\mathrm{GC}/ R_\mathrm{e}$ versus the number of GCs for systems with more than 3 GCs. 
In both panels we find a broad scatter, particularly for galaxies with low numbers of GCs. For those with rich GC systems ($N_\mathrm{GC} > 20$) the average half-number radius is $\sim$3 kpc and 
$R_\mathrm{GC}/R_\mathrm{e}$ ratios is ${\sim}1.2$. Many galaxies have a similar size ratio to that found by \cite{vandokkum2017} for DF44 and DFX1 in the Coma cluster. 
See \cite{forbes2024} for a more detailed discussion of Coma cluster UDGs and the role of the GC system size 
in determining the total number of GCs.  

\subsection{GC System Colours}

We computed mean GC colours after background contaminants were statistically
removed using the same background removal process described above to measure
GC system sizes.
Here too the process was repeated 25 times for each galaxy, and we require a
minimum of 4 GC candidates (after background removal) to measure a mean
colour.
The adopted mean GC colour is the average value of the 25 sample means, with
the uncertainty being the standard deviation of the means.
The mean GC colours are listed in the sixth column of Table \ref{tab:GCs}.

\begin{figure}
    \includegraphics[width=0.485\textwidth]{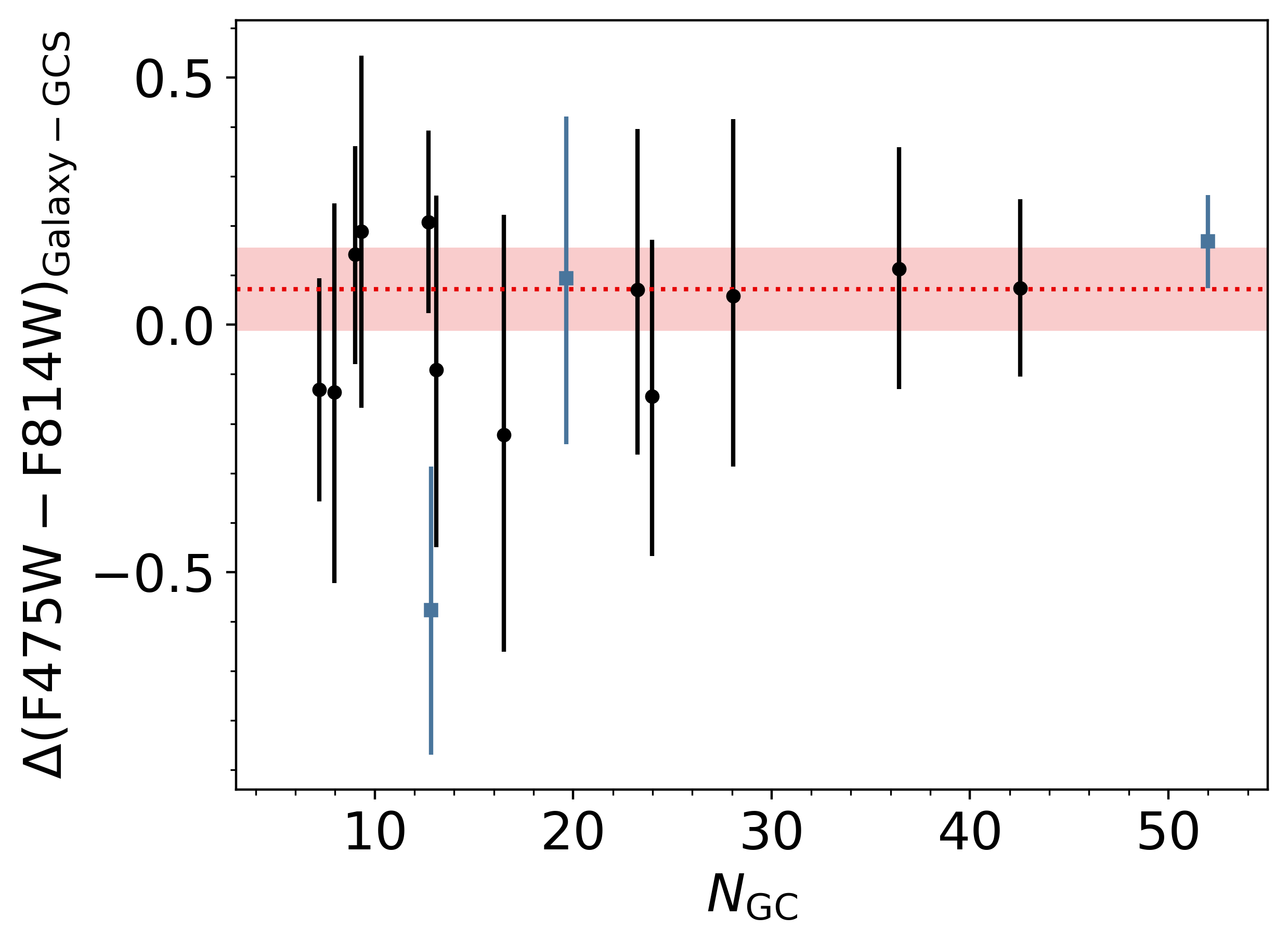}
    \caption{
    Colour difference between host galaxy and the mean colour of its GC system 
    plotted against GC number.
    Black circles are galaxies which satisfy the UDG criteria, blue squares are dwarfs. 
    Positive values indicate galaxies that are redder than their GCs.
    The dotted red line and shaded area are the weighted mean colour
    difference and the weighted error of the mean, respectively, equal to
    $\Delta(\mathrm{F475W} - \mathrm{F814W})_{\mathrm{UDG} - \mathrm{GCS}} =
    0.07 \pm 0.08$ mag.
    Within the uncertainty, galaxies and GC systems have the same mean colour
    indicating the same age and metallicity for their stars.
    \label{fig:colourdiff}
    }
\end{figure}

In Figure \ref{fig:colourdiff} we show the difference between the colour of each galaxy and the mean colour of its GC system.
A positive colour differential value signifies that the galaxy is redder than its GCs (as found for DF2 and DF4 in the NGC 1052 group by \citealt{vandokkum2022}), indicating a more extended star formation history
with metal enrichment. 
The dotted red line and red shaded area show the weighted mean colour
difference and the weighted error of the mean, respectively.
We find a mean colour difference of $\Delta(\mathrm{F475W} -
\mathrm{F814W})_{\mathrm{UDG} - \mathrm{GCS}} = 0.07 \pm 0.08$ mag. Thus we find a slight
positive difference but statistically they have the same colour in the mean. 
Furthermore, each individual galaxy, except for one outlier, has GC colours consistent with their host galaxy.
In a detailed study of the field UDG DGSAT~I, \cite{janssens2022} found
identical colours for the GCs and underlying galaxy light.
\cite{2022MNRAS.511.4633S} also compared the $\mathrm{F475W} - \mathrm{F814W}$
colours of six Coma cluster UDGs with their GC systems finding similar
colours.
These observations are compatible with a scenario whereby the galaxies are
composed of the \emph{same} stellar populations (a combination of age and
metallicity) as their GCs.

Very few UDGs have spectroscopic age and metallicity measures for both the
galaxy stars and the GCs. One such case is NGC~5846\_UDG1 (which is discussed
in detail by \citealt{forbes2024}). Using MUSE on the VLT,
\cite{2020A&A...640A.106M} found the same ages and metallicities, within the
uncertainties, for the galaxy stars and a dozen of the brightest GCs.
Another is NGC~1052-DF2, where \cite{fensch2019} found the same ages, but GCs
with lower metallicity than the galaxy stars.
As noted in the Introduction, UDGs made up of stars that have the same mean
ages and metallicities as their GCs are good candidates for `failed galaxies'.
In this scenario, early quenching of the galaxy stars prevents extended star
formation. Over time GCs are disrupted and their stars contribute, and perhaps
dominate over, the original galaxy stars.

\subsection{Nucleation}

A quick assessment for the presence of stellar nuclei was performed by eye and the results from this are listed in the last column of Table \ref{tab:GCs}. 
In some cases, the same central object is classified as a GC candidate in our automated detection process as well as a stellar nucleus in our visual inspection. 
This highlights the arbitrary nature of determining whether a near-centrally located object is a \textit{bona fide} GC or a true nucleus (which itself may have been formed by the merger of GCs). 
We find 10 of our sample galaxies possess a clear stellar nucleus, a frequency of about 20\%.
This fraction is similar to that found by 
\cite{lim2018} for UDGs in the outer regions of the Coma cluster, and about half the frequency of UDGs located in the Coma cluster core.  
Selection effects need to be taken into account for a detailed comparison. 

We have previously postulated \citep{janssens2019} that the destruction of nucleated UDGs in clusters may be an important
pathway towards the formation of ultra-compact dwarfs (UCDs). UCDs have sizes and luminosities similar to those of stellar nuclei (which may themselves be the merged product of several GCs). The disruption  of the surrounding galaxy by tidal forces is expected to leave only the dense nucleus, which is then identified as a UCD \citep{2013MNRAS.433.1997P}. 
This process has recently received  confirmation by observation of some transition objects between UDGs and UCDs 
in the Virgo cluster \citep{wang2023}.

\section{Summary and Conclusions}

As part of the PIPER survey of the Perseus cluster, we have conducted an analysis of an initial sample of 50 low surface brightness (LSB) galaxies identified in ground-based imaging. Our new data from the {\it Hubble Space Telescope} consists of F475W and F814W images with the ACS camera, along with F475X and F814W images with the WFC3 camera covering all 50 galaxies. We  were able to measure the structural parameters and colours for 41 galaxies, finding that 23 meet the traditional criteria for a UDG. The remaining 18 galaxies have UDG-like properties, having either a slightly brighter surface brightness and/or a more compact size than the standard definition. 
As well as measuring surface brightnesses and half-light radii, we measure total magnitudes, ellipticities, S\'{e}rsic  indices and mean colours for the galaxies.

We identify globular cluster (GC) candidates associated with each galaxy. After corrections for background contamination and magnitude incompleteness, we estimate the total number of GCs hosted by each galaxy. We also measure their mean colour and their spatial extent as quantified by the radius enclosing half of the total number. 

Although most galaxies have few, if any, GCs we find some to reveal GC systems with tens of GCs. These GC-rich UDGs rival those seen in the Coma cluster. Given the general lack of rich GC systems in other clusters (Fornax, Virgo, Hydra~I) this may suggest that only the most massive
($M_{200} \sim 10^{15}~M_{\odot}$) clusters host GC-rich UDGs. 
We also find that the GC system mass to galaxy stellar mass ratio is similar for Perseus cluster LSB galaxies compared to those in the Coma cluster.

We find that the GC systems of our Perseus LSB galaxies with rich GC systems ($N_\mathrm{GC} > 20$) have an extent that is typically around 1.2 times that of their host galaxy half-light radius. This is similar to that found for  well-studied GC rich UDGs in the Coma cluster by \cite{vandokkum2017}. 

The mean colour of our GC systems is the same as that of the host galaxies
within our measurement uncertainty, i.e.\ $0.07 \pm 0.08$ mag.
This suggests that GCs and galaxy stars may have formed at the same epoch from the same enriched
gas. This would be expected in the `failed galaxy' formation scenario for UDGs where star formation is quenched early and subsequent disruption of GCs makes a significant contribution to the 
stellar component of the host UDG.

\section*{Acknowledgements}

We thank the anonymous referee for suggestions which improved the quality of
the manuscript, and B. Koribalski, L. Buzzo and L. Haacke for useful
discussions.
This research is based on observations made with the NASA/ESA Hubble Space
Telescope for program GO-15235 and obtained at the Space Telescope Science
Institute (STScI).
STScI is operated by the Association of Universities for Research in
Astronomy, Inc., under NASA contract NAS 5-26555.
Support for this program was provided by NASA through grants to A.J.R.\ and P.R.D.\ from STScI.
A.J.R.\ was supported by National Science Foundation grant AST-2308390.
This research made use of Astropy,\footnote{\url{http://www.astropy.org}} a
community-developed core Python package for Astronomy \citep{astropy:2013,
astropy:2018}.
This research made use of Photutils, an Astropy package for detection and
photometry of astronomical sources \citep{bradley2021}.
This research has made use of the NASA/IPAC Extragalactic Database (NED),
which is operated by the Jet Propulsion Laboratory, California Institute of
Technology, under contract with the National Aeronautics and Space
Administration.
This research has made use of NASA’s Astrophysics Data System Bibliographic
Services.

\section*{Data Availability}

The raw \textit{HST} data can be obtained from MAST.
For other data, reasonable requests will be considered by the first author.



\bibliographystyle{mnras}
\bibliography{ms} 



\appendix

\section{Completeness Curves}\label{sec:completeness_curves}

\counterwithin{figure}{section}
\setcounter{figure}{0} 

\begin{figure*}
    \includegraphics[width=0.99\textwidth]{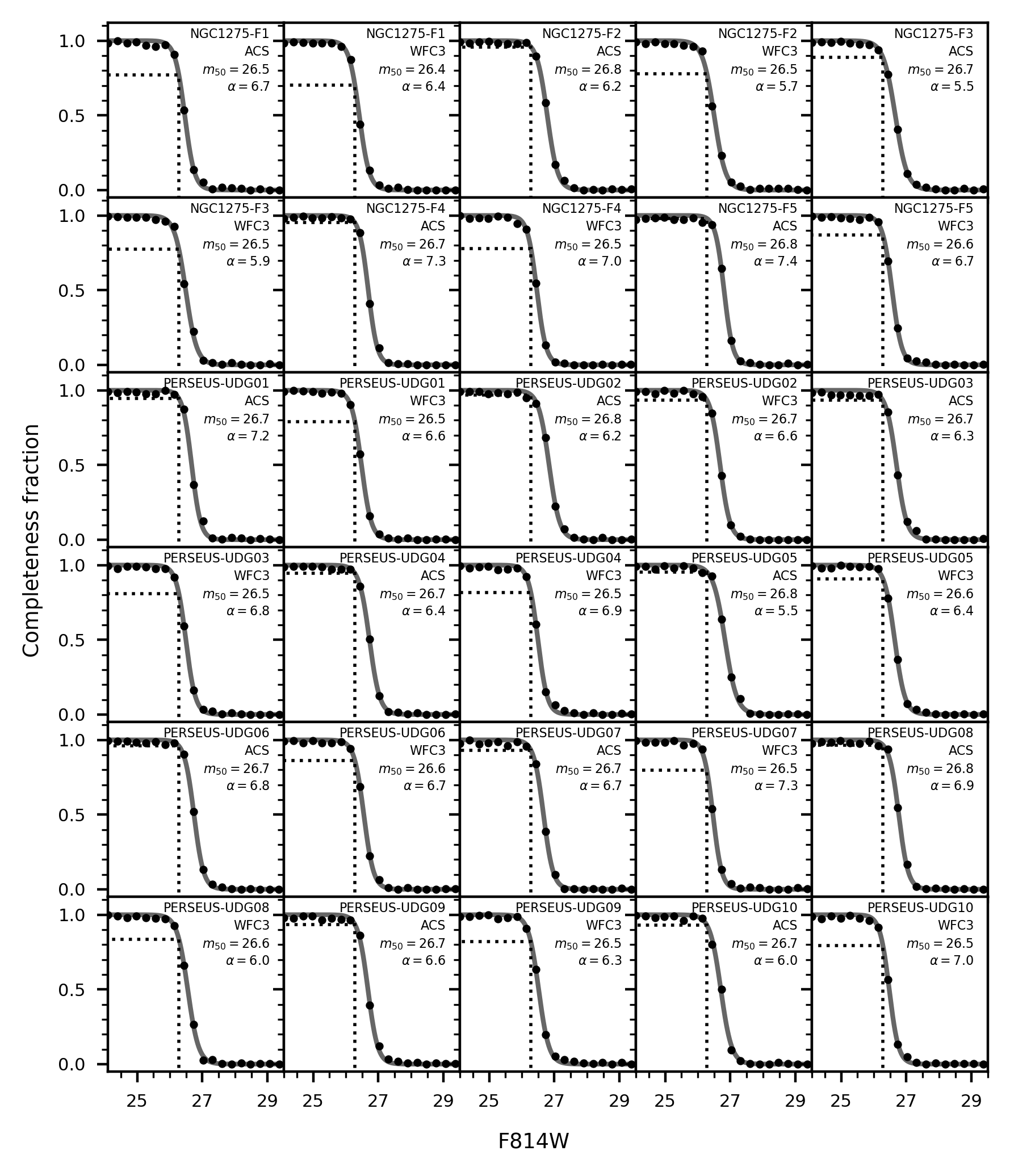}
	\caption{
    F814W completeness curves for all 30 fields (15 visits with
    parallel ACS and WFC3 pointings) as a function of input artificial star magnitude.
    The points are the fraction of artificial stars recovered in bins of size 0.25 magnitudes.
    The solid line is the best fitting function following Equation
    \ref{eqn:fm}.
    Inset text in each panel lists the visit name, camera, 50\% completeness
    limit $m_{50}$ and $\alpha$.
    The dotted line marks the completeness level at the predicted turnover
    magnitude of the GCLF ($m_\mathrm{TO,F814W}$ = 26.3). 
    \label{fig:completeness_curves}
	}
\end{figure*}

In Figure \ref{fig:completeness_curves}, we show the F814W artificial
star completeness curves for all 30 fields.

\bsp	
\label{lastpage}
\end{document}